\documentclass[10pt, letter, onecolumn]{arxiv}
\usepackage{amsmath,amssymb,amsfonts}%
\usepackage{amsthm}%
\usepackage{mathrsfs}%
\usepackage[title]{appendix}%
\usepackage{xcolor}%
\usepackage{textcomp}%
\usepackage{manyfoot}%
\usepackage{booktabs}%
\usepackage{algorithm}%
\usepackage{algpseudocode}
\usepackage{algorithmicx}%
\usepackage{listings}%
\usepackage{pifont}
\usepackage{utfsym}

\usepackage{makecell}
\usepackage{kantlipsum, lipsum}
\usepackage{dm-colors}
\usepackage{pstricks, pst-node}
\usepackage{verbatim}
\usepackage{multirow}
\usepackage{scalerel}
\usepackage{booktabs}
\usepackage{enumitem}
\usepackage{xspace}
\usepackage{bm}
\usepackage{bbm}
\usepackage{mathtools}
\usepackage{soul}
\usepackage{epsfig}
\usepackage{graphicx}
\usepackage{csquotes}
\usepackage{setspace}
\usepackage{colortbl}
\usepackage{threeparttable}
\usepackage{tabularx,ragged2e}
\usepackage{placeins}
\usepackage[hang,flushmargin,symbol]{footmisc}

\usepackage{varioref}
\usepackage[pagebackref=false,breaklinks=false,%
            colorlinks=true,bookmarks=true,citecolor=ourdarkblue,%
            urlcolor=ourdarkblue,linkcolor=ourdarkblue]{hyperref}
\usepackage[noabbrev,capitalize]{cleveref}
\usepackage{etoc}
\usepackage{lineno}
\usepackage{mdframed}

\def\OURMODEL{AutoRG-Brain}
\def\OURDATASET{SSPH}
\def\RELEASERG{RadGenome-Brain MRI}

\def\OURDATASETWOS{SSPH}
\def\RELEASERGWOS{RadGenome-Brain MRI}
\def\OURMODELWOS{AutoRG-Brain}

\definecolor{mygray}{rgb}{0.85, 0.85, 0.85}
\newcommand{\cmark}{\ding{51}}%
\newcommand{\xmark}{\ding{55}}%

\title{\hspace{3pt}\Large{AutoRG-Brain:~Grounded Report Generation \\ for Brain MRI}}

\author{
  Jiayu Lei\textsuperscript{1,2},
  Xiaoman Zhang\textsuperscript{2,3},
  Chaoyi Wu\textsuperscript{2,3},
  Lisong Dai\textsuperscript{4,3},\\
  \normalsize Ya Zhang\textsuperscript{2,3},
  Yanyong Zhang\textsuperscript{1}, 
  Yanfeng Wang\textsuperscript{2,3,$\ast$},
  Weidi Xie\textsuperscript{2,3,$\ast$},
  Yuehua Li\textsuperscript{4,3,$\ast$} \\
   \normalsize \textsuperscript{1}University of Science and Technology of China \\
   \normalsize \textsuperscript{2}Shanghai AI Laboratory  \hspace{8pt} 
   \normalsize \textsuperscript{3}Shanghai Jiao Tong University \\
   \normalsize \textsuperscript{4}Shanghai Sixth People’s Hospital Affiliated to Shanghai Jiao Tong University \\
}

\renewcommand{\correspondingauthor}[1]{$\ast$~Corresponding author. Email addresses: $\{$weidi$\}$@sjtu.edu.cn }

\begin{document}


\begin{abstract}

Radiologists are tasked with interpreting a large number of images in a daily base, with the responsibility of generating corresponding reports. This demanding workload elevates the risk of human error, potentially leading to treatment delays, increased healthcare costs, revenue loss, and operational inefficiencies. 
To address these challenges, 
we initiate a series of work on grounded Automatic Report Generation~(\textbf{AutoRG}), starting from the brain MRI interpretation system, which supports the delineation of brain structures, the localization of anomalies, and the generation of well-organized findings. We make contributions from the following aspects, 
first, on dataset construction, we release a comprehensive dataset encompassing segmentation masks of anomaly regions and manually authored reports, termed as \textbf{\RELEASERGWOS}. This data resource is intended to catalyze ongoing research and development in the field of AI-assisted report generation systems. 
Second, on system design, we propose \textbf{\OURMODELWOS}, the first brain MRI report generation system with pixel-level grounded visual clues. 
Third, for evaluation, we conduct quantitative assessments and human evaluations of brain structure segmentation, anomaly localization, and report generation tasks to provide evidence of its reliability and accuracy. This system has been integrated into real clinical scenarios, where radiologists were instructed to write reports based on our generated findings and anomaly segmentation masks.
The results demonstrate that our system enhances the report-writing skills of junior doctors, aligning their performance more closely with senior doctors, thereby boosting overall productivity.

\end{abstract}

\maketitle
\section{Introduction}

Radiology imaging, serving as a critical disease diagnostic tool, 
plays a crucial role in clinical practice. Despite their importance, writing radiological reports is time-consuming and prone to errors due to the heavy workloads of radiologists~\cite{mcdonald2015effects}, which can lead to significant healthcare delays and risks, particularly with junior clinicians~\cite{giles2019error}. Therefore, an ideal radiology image interpretation system should enable to enhance lesion localization, 
improve report writing efficiency and educate less experienced clinicians. 

In the recent literature, automated report generation has shown significant progress with the development of visual-language models~\cite{liu2019clinically, li2023dynamic, nooralahzadeh2021progressive, zhou2021visual, tu2024towards, wu2023towards, chen2020generating}, 
however, most studies have predominantly focused on chest X-rays, 
driven by the prevalence of large-scale image-text datasets specific to this modality~\cite{johnson2019mimic,demner2016preparing}.
This is far from meeting the demand of clinical practices where other imaging modalities, like MRI or CT, are also critical for accurate diagnosis. 
Additionally, existing methods directly generate reports based on entire scans, without adequately addressing specific anomalies or anatomical regions, which may potentially lead to omissions in detailed regional descriptions, such as fine-grained analyses of crucial lesions or abnormal regions, consequently limiting the interpretive value and usability of these report generation systems by physicians.

In response to the above-mentioned limitations, 
we propose a novel system that decomposes the original report generation task into two stages, namely, an anomaly regions of interest~(ROI) generation stage and a visual prompting guided report generation stage. 
As a proof of concept, we focus on brain MRI report generation to demonstrate the superiority of our system, and make contribution from three aspects, namely, dataset construction, model design and training, and lastly, comprehensive evaluations.

Specifically, on \textbf{dataset construction}, 
we collect a large-scale dataset for pre-training, 
which encompasses an in-house clinical dataset with 30,793 multi-modal scans and reports from Shanghai Sixth People’s Hospital~(\textbf{\OURDATASETWOS}). Additionally, we curate a dataset for grounded report generation, termed as \textbf{RadGenome-Brain}
\begin{figure}[!ht]
\centering
\includegraphics[width=\textwidth]{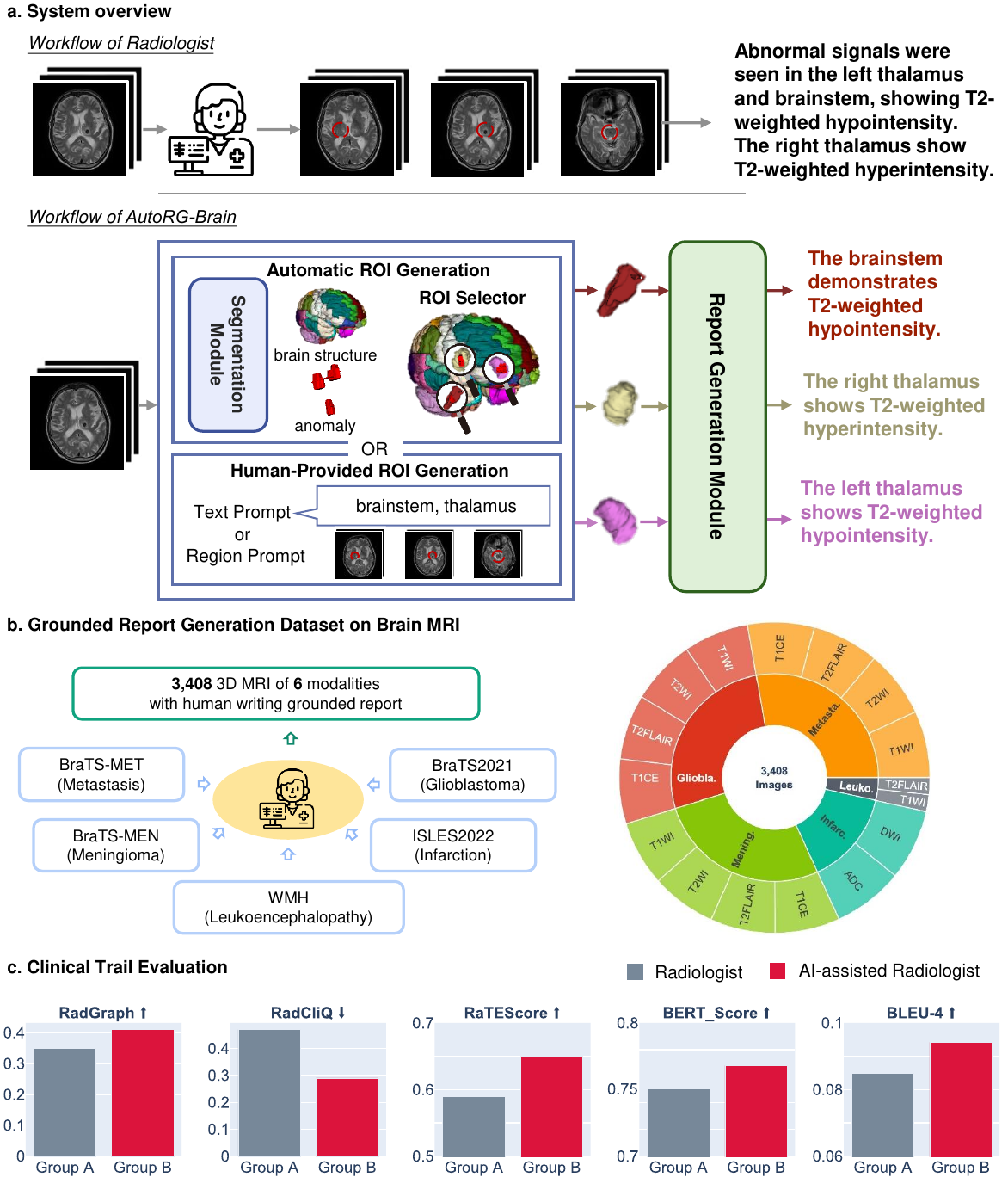}
\caption{Overview of our contributions. 
\textbf{a} An analogy between the radiologists and the pipeline of our proposed system \textbf{\OURMODEL} on report generation.  
\textbf{b} The proposed grounded report generation dataset (\textbf{\RELEASERG}), contains 3,408 region-report pairs, covering five diseases and six MRI modalities. \textbf{c} The result of \textbf{\OURMODEL} advancing radiologist report writing efficiency, making the report quality closer to the gold standard report written by senior doctors.
}
\label{fig:system_overview}
\end{figure}
\clearpage

\textbf{MRI}\footnote{This is part of the RadGenome dataset series, for example, RadGenome-Chest CT~\cite{zhang2024radgenome-chest-ct}, RadGenome-Brain MRI, with the goal of building large-scale multi-modal radiology datasets, with global reports, structure segmentation, abnormal segmentation, and regional reports.}, with 3,408 multi-modal scans, reports, and ground truth anomaly segmentation masks, compiled from the publicly available datasets, for example, BraTS2021~\cite{bakas2017segmentation,bakas2017advancing,menze2014multimodal,baid2021rsna}, BraTS-MEN~\cite{labella2023asnr}, BraTS-MET~\cite{moawad2023brain}, ISLES2022~\cite{hernandez2022isles} and WMH~\cite{kuijf2019standardized}, 
covering 6 modalities: T1-weighted, T2-weighted, DWI, T2-Flair, ADC, 
and T1-contrast. For each patient case, we ask the radiologists to write findings and impressions for the annotated anomaly regions. 
We will release \textbf{RadGenome-Brain MRI} to the community, 
supporting future model training and evaluation.

On system design, we propose \textbf{\OURMODELWOS}, 
the first regional brain MRI report generation system, 
that enables comprehensive segmentation of each anomaly region and generation of well-organized narratives, 
to describe observations in different anatomical regions as shown in Fig.~\ref{fig:system_overview}. 
In more detail, \textbf{\OURMODEL} consists of two components, namely an automatic ROI generation component, and a visual prompting guided report generation component. The former component outputs a regional mask for each detected anomaly, serving as auto-initialized visual prompts for the latter report generation module, enabling the regional-related reports accordingly.

On evaluation, we conducted comprehensive experiments to assess the performance of \textbf{\OURMODEL} on brain structure segmentation, anomaly segmentation, and report generation. 
Both automatic metrics and human rating scores validate the model's superior generalizability and effectiveness. 
Subsequently, we integrated our model into the \textbf{real clinical scenarios} to assess its potential applicability, 
where we have developed an online platform for radiologists to compose textual responses based on provided 2D/3D images and textual instructions. Two groups of junior radiologists~(2-5 years) are instructed to produce reports based on a commonly used template or our generated reports and anomaly masks respectively. The results demonstrate that, with the assistance of our system, junior radiologists are also able to  discover subtle abnormalities, improving the quality of final report writing. 
We will publicly release our codes, trained models, annotated datasets, and the platform for evaluation to support further research.
\section{Results}
In this section, we start by introducing our problem scenarios,
and then present the evaluation results of two key sub-modules of our system: the segmentation module included in the automatic ROI generation component and the visual prompting guided report generation module. 
Lastly, we showcase the efficiency of integrating \OURMODEL~into a real clinical scenario.

\subsection{Problem Scenario}
\label{problem scenarios}

This paper aims to develop a comprehensive Brain MRI interpretation system that provides paired visual grounding results and regional radiology reports for clinicians, which allows the clinicians to efficiently review and revise generated reports based on visual clues, significantly improving report writing efficiency. 
Specifically, we decompose this task into two different stages, \emph{i.e.}, an automatic ROI generation stage and a visual prompting guided report generation stage.

Formally, given a brain MRI scan $X \in \mathbb{R}^{H\times W\times D}$ in any modality, {\em e.g.}, T1, T2, Flair, where $H$, $W$ and $D$ denote the height, width, and depth respectively, the automatic ROI generation module first generate two global segmentation masks using the segmentation module ($\Phi_{\text{SEG}}$), as:
\begin{equation}
    \{\hat{s}_a, \hat{s}_b\} =  \Phi_{\text{SEG}}(X),
\end{equation}
$\hat{s}_a, \hat{s}_b$ denote the predicted global anomaly and brain structure masks respectively. From the brain structure segmentation labels, an ROI selector derives $\hat{s}_b = \{b_1, b_2, \dots b_m\}$, 
with each $b_i$ being a mask for a certain brain structure region, 
for example, brainstem, left temporal lobe, right thalamus, etc.
It also divides the binary global anomaly segmentation mask into a sequence of connected components, {\em i.e.}, $\hat{s}_a =\{ a_1, a_2, \dots, a_n\}$. Then, for each $a_{i}$, the ROI selector finds a group of brain regions overlaid by $a_{i}$ to form a regional mask prompt $m_{i}$,
{\em i.e.}, $m_i = \bigcup\limits_{ b_j \in \hat{s}_b,  b_j \cap a_i \neq \emptyset } b_j$, for subsequent regional report generation.
Notably, this step can also be replaced by human interactions where users can directly highlight their desired ROIs by providing a group of brain structure names or segmentation masks.

Then, we use our region-guided report generation module ($\Phi_{\text{RG}}$) to describe each $m_i$, as:
\begin{equation}
    \hat{T}_i = \Phi_{\text{RG}}(X,m_i),
\end{equation}
where $\hat{T}_i$ is the predicted regional radiology report corresponding to the human-provided or auto-segmented visual prompt $m_i$.
By combining all regional reports and completing the left normal regions with standard templates, we can thus get a predicted global report. The illustration of the above steps is shown in Fig.~\ref{fig:system_overview}a.

\subsection{Evaluation on Segmentation}\label{sec:sub-module_eval}
In this section, we present the results on brain anomaly and brain structure segmentation, as shown in Tab.~\ref{tab:anomaly_seg_ZS}. 
To our knowledge, our segmentation module is the first model that enables to segment both abnormalities and structures for multi-modal brain MRIs. Specifically, the training undergoes two stages, with the first self-supervised training stage on \OURDATASETWOS, and the second semi-supervised training stage on \OURDATASET~and 9 public datasets, with more details in Sec.~\ref{sec:method}. 
We start with the quantitative results for both tasks, 
comparing the performance at different stages with baselines. 
Then, we present the human evaluation on our second-stage segmentation module for brain structure segmentation on abnormal MRI scans.

\begin{table}[!htb]
\centering
\caption{Comparison of \OURMODEL~with SOTA unsupervised brain MRI segmentation models on public datasets, BraTS2021, ISLES2022, and Hammers-n30r95. The data for the benchmark method comes from paper~\cite{zhang2023unsupervised}. Ours-S1 is our segmentation module after the first self-supervised training stage and Ours-S2 is our segmentation module after the second semi-supervised training stage. We report the Dice Similarity Coefficient (DSC), Precision (PRE) and Sensitivity (SE) scores. The best result is bolded, and the second best result is underlined.}
\vspace{3pt}
\renewcommand{\arraystretch}{1.1}
\footnotesize
\resizebox{1.0\linewidth}{!}{
\begin{tabular}{lcccccc}
\toprule
 \rowcolor{mygray} \multicolumn{7}{c}{Brain Anomaly Segmentation} \\
 \midrule
\multirow{2}{*}{Method} &\multicolumn{3}{c}{BraTS2021} &\multicolumn{3}{c}{ISLES2021} \\
\cmidrule(lr){2-4} \cmidrule(lr){5-7}
  & DSC (\%) $\uparrow$ & PRE (\%) $\uparrow$ & SE (\%) $\uparrow$ & DSC (\%) $\uparrow$ & PRE (\%) $\uparrow$ & SE (\%) $\uparrow$ \\
\midrule
AE~\cite{ballard1987modular} & 33.64\ $\pm$\ 21.45 & 62.09\ $\pm$\ 30.46 & 30.00\ $\pm$\ 19.37 & 18.00\ $\pm$\ 21.45 & 38.06\ $\pm$\ 39.93 & 15.06\ $\pm$\ 15.06 \\
VAE~\cite{kingma2013auto} & 42.67\ $\pm$\ 18.30 & 56.53\ $\pm$\ 25.18 & 39.63\ $\pm$\ 19.66 & 13.10\ $\pm$\ 19.39 & \underline{45.84\ $\pm$\ 48.38} & 15.06\ $\pm$\ 15.06 \\
VQ-VAE~\cite{van2017neural} & 31.62\ $\pm$\ 26.24 & 74.38\ $\pm$\ 34.02 & 30.98\ $\pm$\ 29.73 & 16.82\ $\pm$\ 23.49 & 31.59\ $\pm$\ 36.23 & 14.18\ $\pm$\ 20.95 \\
f-AnoGAN~\cite{schlegl2019f} & 39.36\ $\pm$\ 19.00 & 36.65\ $\pm$\ 22.94 & 53.51\ $\pm$\ 18.48 & 17.18\ $\pm$\ 20.23 & 28.54\ $\pm$\ 35.39 & 18.38\ $\pm$\ 21.16\\
DCTAE~\cite{ghorbel2022transformer} & 38.65\ $\pm$\ 29.06 & 74.40\ $\pm$\ 34.11 & 40.56\ $\pm$\ 34.46 & 16.98\ $\pm$\ 23.48 & 30.29\ $\pm$\ 34.96 & 14.65\ $\pm$\ 21.36 \\
BTAE~\cite{ghorbel2022transformer} & 30.54\ $\pm$\ 30.92 & 64.04\ $\pm$\ 41.71 & 32.86\ $\pm$\ 35.76 & 18.72\ $\pm$\ 25.02 & 31.07\ $\pm$\ 35.14 & 16.73\ $\pm$\ 23.41\\
HTAES~\cite{ghorbel2022transformer} & 14.57\ $\pm$\ 24.07 & 35.36\ $\pm$\ 44.04 & 15.37\ $\pm$\ 27.48 & 14.26\ $\pm$\ 22.66 & 32.20\ $\pm$\ 38.19 & 11.35\ $\pm$\ 19.63 \\
CUAS~\cite{silva2022constrained} & 54.74\ $\pm$\ 18.57 & 74.16\ $\pm$\ 25.12 & 47.71\ $\pm$\ 19.23 & 21.31\ $\pm$\ 22.66 & 21.11\ $\pm$\ 25.12 & 42.20\ $\pm$\ 34.71 \\
3D IF~\cite{naval2021implicit} &  47.10\ $\pm$\ 17.21 & 54.51\ $\pm$\ 24.54 & 48.79\ $\pm$\ 16.52 & 14.28\ $\pm$\ 19.79 & 17.37\ $\pm$\ 29.74 & 35.56\ $\pm$\ 27.97\\
Sim2Real~\cite{zhang2023self} &  66.44\ $\pm$\ 22.44 & 62.32\ $\pm$\ 26.68 & \underline{81.97\ $\pm$\ 20.97} & 23.82\ $\pm$\ 25.08* & 18.77\ $\pm$\ 24.53 & \underline{65.57\ $\pm$\ 31.87}\\
\midrule
Ours-S1 & \underline{75.69\ $\pm$\ 16.49} & \underline{82.50\ $\pm$\ 12.30} & 74.38\ $\pm$\ 22.19 & \underline{38.66\ $\pm$\ 31.24} &  44.66\ $\pm$\ 36.49 & 42.80\ $\pm$\ 31.46\\
\midrule
Ours-S2 & \textbf{90.10\ $\pm$\ 7.64 }& \textbf{92.40\ $\pm$\ 7.39} & \textbf{88.75\ $\pm$\ 10.37} & \textbf{71.14\ $\pm$\ 22.81} &  \textbf{70.48\ $\pm$\ 22.38} & \textbf{77.71\ $\pm$\ 24.85}\\
\midrule
 \rowcolor{mygray} \multicolumn{7}{c}{Brain Structure Segmentation} \\
 \midrule
\multirow{2}{*}{Method} &\multicolumn{3}{c}{Hammers-n30r95} & \multicolumn{3}{c}{\makecell{Hammers-n30r95 \\ w. Synthetic Anomalies}} \\
\cmidrule(lr){2-4} \cmidrule(lr){5-7}
  & DSC (\%) $\uparrow$ & PRE (\%) $\uparrow$ & SE (\%) $\uparrow$ & DSC (\%) $\uparrow$ & PRE (\%) $\uparrow$ & SE (\%) $\uparrow$ \\
\midrule
Ours-S1 & \underline{67.43$\pm$1.04} & \underline{64.31$\pm$0.72} & \underline{66.35$\pm$1.74} & \underline{66.19$\pm$1.44} & \underline{63.59$\pm$1.02} & \underline{65.18$\pm$1.93} \\
\midrule
Ours-S2 & \textbf{68.78\ $\pm$\ 0.79} & \textbf{68.80\ $\pm$\ 0.87} & \textbf{71.33\ $\pm$\ 1.22} & \textbf{67.60\ $\pm$\ 0.62} &  \textbf{67.62\ $\pm$\ 0.29} & \textbf{70.14\ $\pm$\ 1.14}\\
\bottomrule
\end{tabular}}
\label{tab:anomaly_seg_ZS}
\end{table} 

\noindent \textbf{Brain Anomaly Segmentation.} 
We compare both our first-stage segmentation module (Ours-S1) and second-stage segmentation module (Ours-S2) with existing segmentation models on BraTS2021 and ISLES2022, and report Dice Similarity Coefficient (DSC), Precision (PRE) and Sensitivity (SE).  
As shown in Tab.~\ref{tab:anomaly_seg_ZS}, Ours-S1 exhibits competitive performance, surpassing other state-of-the-art (SOTA) unsupervised models by noticeable, consistent margins on DSC across both datasets, achieving a DSC of 75.69\% on BraTS2021 and a DSC of 38.66\% on ISLES2022. Although Sim2Real achieves the highest sensitivity among unsupervised methods on both datasets, its precision is significantly lower, indicating that there may be numerous false positive areas in its segmentation results, as demonstrated in the qualitative comparison results presented in Appendix Fig.~\ref{fig:anomaly_segmentation_example}. After the second training stage, our model~(Ours-S2) further enhances the anomaly segmentation performance, resulting in much higher performance on both datasets.

\noindent \textbf{Brain Structure Segmentation.} 
Existing methods typically focus on segmenting structures in healthy MRI brains~\cite{zhao2023model}. In contrast, our segmentation module is designed to handle MRI scans of lesion-induced brains, which may exhibit regional expansion, compression, or deformation. We evaluate the performance of our brain structure segmentation on the public dataset Hammers-n30r95. Due to the lack of existing datasets with ground truth structure segmentation masks for multi-modal abnormal brain MRIs, we generated a synthetic test set by adding artificial `lesions' to the original scans~\cite{zhang2023self}, 
denoted as `Hammers-n30r95 with Synthetic Anomalies'. 
Our method, Ours-S1, demonstrates robustness on abnormal brains, achieving a Dice Similarity Coefficient (DSC) score of 67.43\% on `Hammers-n30r95' compared to 66.19\% on `Hammers-n30r95' with Synthetic Anomalies'. The second semi-supervised stage, Ours-S2, further improves brain structure segmentation performance on both healthy and abnormal brain MRIs, with a DSC score of 68.78\% on `Hammers-n30r95' and 67.60\% on `Hammers-n30r95 with Synthetic Anomalies'.

\begin{figure}[t]
\centering
\includegraphics[width=\textwidth]{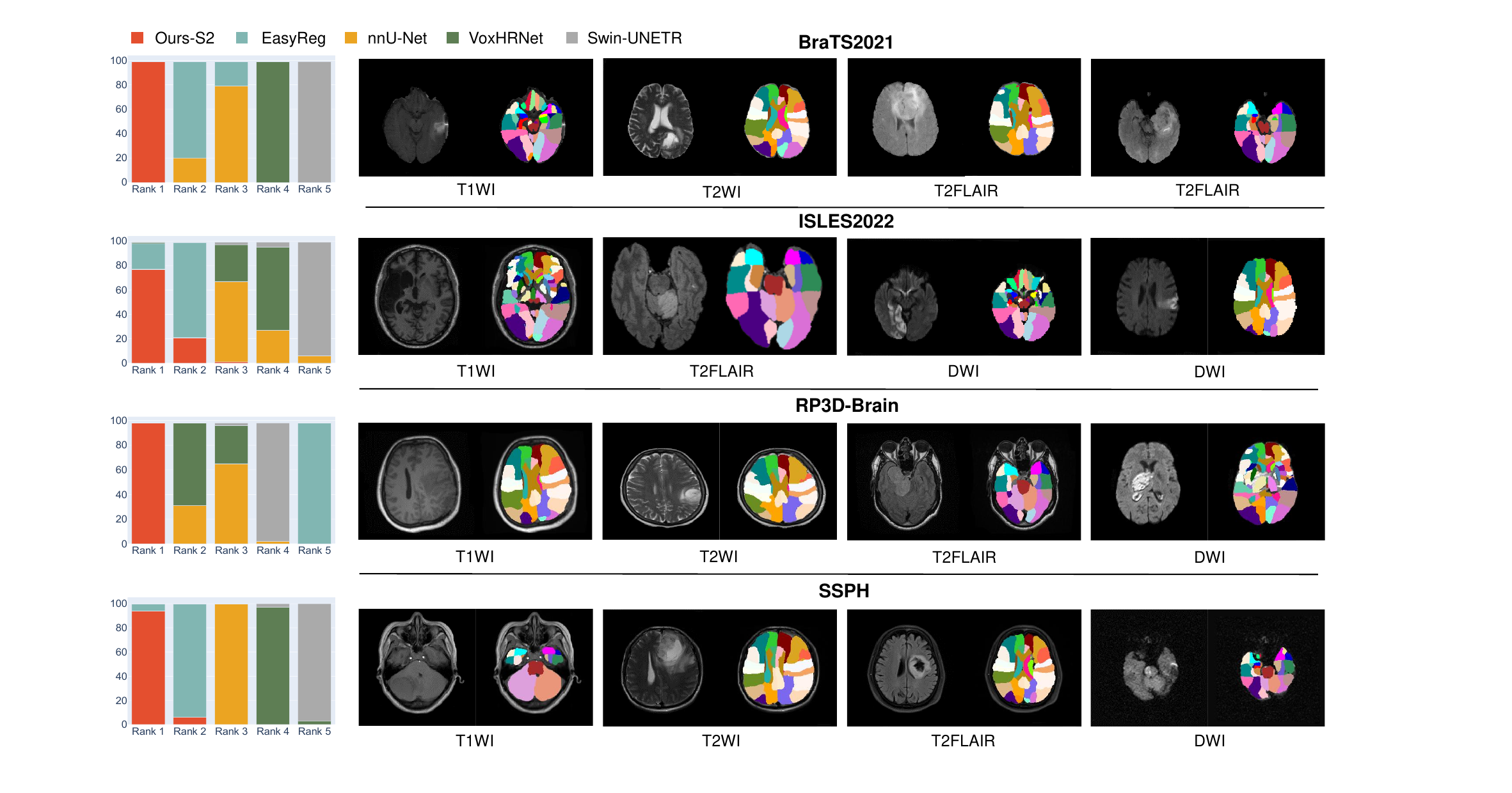}
\caption{
Comparison of our second-stage segmentation module (Ours-S2) with SOTA segmentation backbones and brain registration models on brain structure segmentation for multi-modal brain MRIs with real anomalies. We evaluate these models on four distinct datasets, BraTS2021, ISLES2022, RP3D-Brain, and \OURDATASETWOS. Due to the absence of ground truth brain structure segmentation on those datasets, we present the results based on human rankings on the left, where lower rankings indicate better outcomes. Note the baseline nnU-Net is not the same structure as our segmentation module, since we modify the original nnU-Net to fit our problem scenario.}
\label{anatomy_seg_score}
\end{figure}

To further verify the brain structure segmentation ability of Ours-S2 on multi-modal brain MRIs with real anomalies, we invite \textbf{two radiologists} to rank the predictions from our model~(Ours-S2) and baseline models on scans with various anomalies, derived from ISLES2022, BraTS2021, RP3D-Brain, and \OURDATASETWOS. As shown in Fig.~\ref{anatomy_seg_score}, predictions from Ours-S2 are mostly ranked as the top, and the structure segmentation results of Ours-S 2 are displayed on the right. In general, \OURMODEL~currently demonstrates the best performance on brain structure segmentation for multi-modal lesion-induced brain MRIs across different datasets.

\subsection{Evaluation on Report Generation}\label{sec2_2}  
In this section, we evaluate the system under two settings: 
standard global report generation that describes the entire image,
and grounded report generation, that focuses on generating a report for a given regional visual prompt. Note that, for the global report generation, \OURMODEL~has three inference strategies as introduced in Sec.~\ref{sec:method}, namely, 
(i) \textbf{\OURMODELWOS-Global}: generating global report with global image feature; (ii) \textbf{\OURMODELWOS-AutoSeg}: generating global report by concatenating the grounded reports for auto-segmented regions; 
(iii) \textbf{\OURMODELWOS-Prompt}: generating global report by concatenating the grounded reports with human prompting, {\em i.e.}, specify the region of interest on the structure segmentation results.

\begin{table}[!htb]
\centering
\footnotesize
    \caption{Comparison of \OURMODEL~with SOTA medical foundation models for report generation on \OURDATASET~and \RELEASERGWOS. We divide the verification into verification for generating global reports (\OURDATASET~Global, \RELEASERG~Global) and verification for generating grounded reports (\OURDATASET~Region). We show the results of \OURMODEL~under three test modes on \OURDATASET~Global, namely, (i) \OURMODELWOS-Global. Generate global report using global image feature; (ii) \OURMODELWOS-AutoSeg. Generate global report by concatenating the grounded reports for auto-segmented regions; (iii) \OURMODELWOS-Prompt. Generate global report by concatenating the grounded reports for human-provided regions. `FS' denotes few-shot inference, `ZS' denotes zero-shot inference. BLEU-4 (\%), ROUGE-1 (\%), Bert-Score (\%), RadGraph (\%),  RadCliQ, and RaTEScore (\%) are reported.}
    \vspace{3pt}
    \label{tab:rg_statistic}
    \resizebox{1.0\linewidth}{!}{
   \begin{tabular}{lccc|cccccc}
    \toprule
    Method &Test Mode &Size &  Pretrain Dataset & BLEU-4 $\uparrow$ & ROUGE-1 $\uparrow$ & Bert-Score $\uparrow$ & RadGraph $\uparrow$ & RadCliQ $\downarrow$ & RateScore $\uparrow$ \\ 
    \midrule
    \rowcolor{mygray} \multicolumn{10}{c}{\OURDATASET~Global} \\ 
    \midrule
    MedVInT~\cite{zhang2023pmc} & Global &7B &  PMC-VQA & 0.005$\pm$0.1  & 6.02$\pm$3.45 & 44.52$\pm$5.11 &  1.24$\pm$3.30 & 1.85$\pm$0.31 & 21.19$\pm$20.67\\
    OpenFlamingo ZS~\cite{awadalla2023openflamingo}& Global &9B &  Web\ Data & 0.01$\pm$0.15 & 12.63$\pm$4.28 & 44.31$\pm$3.13 & 0.98$\pm$2.88  & 1.92$\pm$0.17 &29.33$\pm$7.77 \\
    MedFlamingo ZS~\cite{moor2023med}& Global &9B & MTB, PMC-OA & 0.10$\pm$0.39 & 19.56$\pm$5.66 & 51.25$\pm$4.72 & 5.30$\pm$4.80  & 1.75$\pm$0.23 & 39.33$\pm$7.52\\
    \midrule
    OpenFlamingo FS~\cite{awadalla2023openflamingo}& Global &9B &  Web\ Data & 0.60$\pm$1.31 & 32.39$\pm$8.25 & 67.12$\pm$4.09 & 19.26$\pm$6.54 & 1.00$\pm$0.23 & 51.66$\pm$7.90\\
    MedFlamingo FS~\cite{moor2023med}& Global &9B & MTB, PMC-OA & 0.58$\pm$1.38 & 23.30$\pm$7.79 & 65.79$\pm$4.99 & 16.93$\pm$7.33  & 1.13$\pm$0.26 & 46.54$\pm$9.03\\
    \midrule
    \OURMODEL & Global &2B & \OURDATASET & 5.13$\pm$7.75 & 40.88$\pm$16.73 & 65.53$\pm$8.62 & 33.39$\pm$15.58 & 0.90$\pm$0.62 & 60.68$\pm$12.66\\
    \OURMODEL & Global &2B & \RELEASERG+\OURDATASET & 5.95$\pm$8.56 & 40.06$\pm$14.92 & 70.98$\pm$7.00 & 34.48$\pm$14.45 & 0.71$\pm$0.52 & 61.39$\pm$12.23\\
    \midrule
    \OURMODEL & AutoSeg &2B & \OURDATASET & 3.52$\pm$5.59 & 37.62$\pm$17.73 & 69.26$\pm$6.09 &  30.26$\pm$13.68 & 0.66$\pm$0.43&60.49$\pm$11.90  \\
    \OURMODEL & AutoSeg &2B & \RELEASERG+\OURDATASET & 5.94$\pm$7.40 & 44.10$\pm$16.15 & 71.52$\pm$5.95 & 36.31$\pm$13.86 &  0.53$\pm$0.44& 64.74$\pm$11.50\\
    \midrule
    \OURMODEL & Prompt &2B & \OURDATASET & 8.44$\pm$10.45 & 44.44$\pm$14.89 & 72.77$\pm$6.98 & 39.28$\pm$16.32 & 0.39$\pm$0.50 & 66.54$\pm$12.59\\
    \OURMODEL & Prompt &2B & \RELEASERG+\OURDATASET & \textbf{9.93$\pm$11.45} & \textbf{46.76$\pm$14.63} & \textbf{74.35$\pm$6.63} & \textbf{41.74$\pm$15.93} & \textbf{0.30$\pm$0.48} &\textbf{68.76$\pm$12.53} \\
    \midrule
    \rowcolor{mygray} \multicolumn{10}{c}{\OURDATASET~Region} \\ 
    \midrule
    MedVInT~\cite{zhang2023pmc}& Region&7B &  PMC-VQA & 0.04$\pm$0.89 & 11.65$\pm$10.53 & 49.03$\pm$11.00 & 1.50$\pm$5.57 & 1.48$\pm$0.34 & 19.11$\pm$20.00\\
    OpenFlamingo ZS~\cite{awadalla2023openflamingo}&Region &9B &  Web\ Data & 0.006$\pm$0.11 & 13.42$\pm$12.12 & 45.50$\pm$5.62 & 0.00$\pm$0.00 &  1.79$\pm$0.41 &37.76$\pm$13.87 \\
    MedFlamingo ZS~\cite{moor2023med}&Region &9B & MTB, PMC-OA & 0.02$\pm$0.17 & 24.17$\pm$14.17 & 45.00$\pm$4.52 & 6.41$\pm$7.86  &  1.78$\pm$0.31& 39.80$\pm$11.85\\
    \midrule
    OpenFlamingo FS~\cite{awadalla2023openflamingo}&Region &9B & Web\ Data & 0.08$\pm$0.85 & 15.85$\pm$13.19 & 42.41$\pm$8.00 &  2.77$\pm$5.12 & 1.55$\pm$0.35 & 24.16$\pm$18.50\\
    MedFlamingo FS~\cite{moor2023med}&Region &9B & MTB, PMC-OA & 1.85$\pm$4.62 & 31.85$\pm$24.45 & 58.93$\pm$12.51 & 15.64$\pm$22.83  &  1.39$\pm$0.84 & 47.87$\pm$23.27\\
    \midrule
    \OURMODEL & Prompt &2B & \OURDATASET & 7.44$\pm$17.79 & 39.25$\pm$23.31 & 69.61$\pm$12.27 & 29.78$\pm$24.47 & 0.40$\pm$0.67 & 57.07$\pm$21.52\\
    \OURMODEL & Prompt &2B & \RELEASERG+\OURDATASET & \textbf{8.34$\pm$19.23} & \textbf{39.69$\pm$23.47} & \textbf{70.98$\pm$12.09} & \textbf{32.29$\pm$24.18} & \textbf{0.32$\pm$0.65} & \textbf{59.53$\pm$21.12}\\
    \midrule
    \rowcolor{mygray} \multicolumn{10}{c}{\RELEASERG~Global} \\ 
    \midrule
    MedVInT~\cite{zhang2023pmc}& Global &7B &  PMC-VQA & 0.01$\pm$0.30 & 7.30$\pm$3.52 & 48.53$\pm$3.84 & 1.78$\pm$3.94 & 1.56$\pm$0.21 & 34.05$\pm$10.73\\
    OpenFlamingo ZS~\cite{awadalla2023openflamingo}&Global &9B & Web\ Data & 0.03$\pm$0.21 & 14.59$\pm$5.12 & 40.56$\pm$2.89 & 1.98$\pm$5.33 & 1.91$\pm$0.21& 26.55$\pm$9.85\\
    MedFlamingo ZS~\cite{moor2023med}&Global &9B & MTB, PMC-OA & 0.15$\pm$0.50 & 24.91$\pm$8.12 & 48.99$\pm$4.35 & 10.79$\pm$9.44 & 1.57$\pm$0.29 & 37.53$\pm$10.75\\
    \midrule
    OpenFlamingo FS~\cite{awadalla2023openflamingo}&Global &9B &  Web\ Data & 0.25$\pm$1.03 & 30.53$\pm$9.09 & 61.38$\pm$5.68 & 15.91$\pm$9.47 & 1.05$\pm$0.25 & 42.17$\pm$11.51\\
    MedFlamingo FS~\cite{moor2023med}&Global &9B & MTB, PMC-OA & 0.18$\pm$0.91 & 20.26$\pm$6.04 & 57.50$\pm$5.25 & 9.35$\pm$8.64 & 1.24$\pm$0.21 & 36.03$\pm$13.23\\
    \midrule
    \OURMODEL & Prompt &2B & \RELEASERG & \textbf{3.79$\pm$5.45} & \textbf{39.70$\pm$12.59} & \textbf{69.27$\pm$6.48} & \textbf{28.75$\pm$13.89} & \textbf{0.54$\pm$0.37}& \textbf{46.65$\pm$14.02}\\
    \OURMODEL & Prompt &2B & \RELEASERG+\OURDATASET & 3.64$\pm$5.52 & 36.51$\pm$13.53 & 68.01$\pm$6.67 &  26.90$\pm$15.14 &  0.59$\pm$0.40 & 45.95$\pm$15.13\\
    \bottomrule
    \end{tabular}}
\end{table}

\begin{figure}[!htb]
\centering
\includegraphics[width=\textwidth]{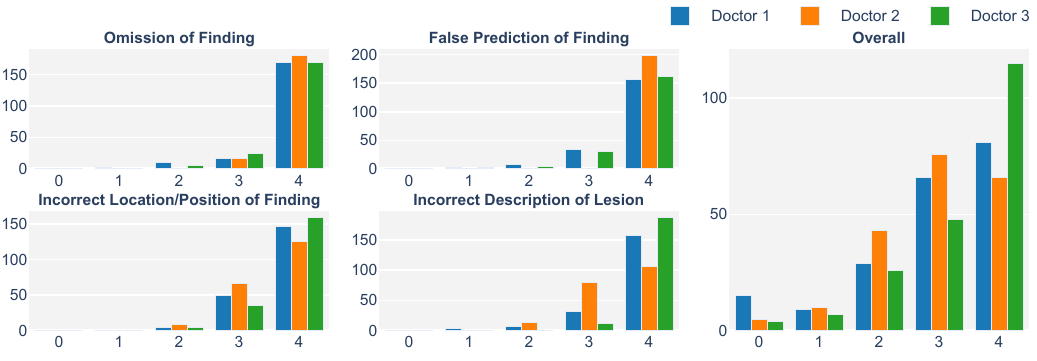}
\caption{The global report generated by \OURMODELWOS-AutoSeg on \OURDATASET~is evaluated by three radiologists in four dimensions: (i) Omission of finding; (ii) False prediction of finding; (iii) Incorrect location/position of finding; (iv) Incorrect description of lesion. 
The scores of four dimensions are shown on the left with the overall score shown on the right. The scoring is on a 4-point scale: 0 indicates more than three clinically significant errors or omissions, 1 indicates three clinically significant errors or omissions, 2 indicates two clinically significant errors or omissions, 3 indicates one clinically significant error or omission, and 4 indicates no clinically significant errors or omissions.}
\label{fig:human_eval_report}
\end{figure}

\noindent \textbf{Global Report Generation.} 
Our proposed system enables report generation with or without human prompting, largely outperforming existing foundational models. 
On \OURDATASET~dataset~(the first part in Tab.~\ref{tab:rg_statistic}), \OURMODELWOS-Prompt trained on both \OURDATASET~and \RELEASERG~reaches the best performance, achieving a RadGraph score of 41.74$\pm$15.93\%, a RadCliQ score of 0.30$\pm$0.48, and a RaTEScore of 68.76$\pm$12.53\%. Additionally, \OURMODELWOS-AutoSeg pre-trained on \OURDATASET~and \RELEASERG~reaches the second best result, which indicates the effectiveness of incorporating the grounded visual clues into report generation. On \RELEASERG~dataset, \OURMODELWOS-Prompt pre-trained on \RELEASERG~outperforms all baselines with a RadGraph score of 28.75$\pm$13.89\%, a RadCliQ score of 0.54$\pm$0.37, and a RaTEScore of 46.65$\pm$14.02\%. The result sets a new benchmark on our proposed open-source brain MRI grounded report generation dataset, which we hope will drive further development of grounded report generation in the research community.

In addition, we conduct a \textbf{human evaluation} on the generated reports~(200 cases selected from SSPH). The annotation page is shown in Appendix Fig.~\ref{fig:human_evaluation_annotation}. We use \OURMODELWOS-AutoSeg as the default setting, three radiologists were invited to rate the quality of the generated report from four aspects~\cite{yu2023evaluating}: 
    (i) Omission of finding; 
    (ii) False prediction of finding; 
    (iii) Incorrect location/position of finding; 
    (iv) Incorrect description of lesion. 
Each of the above dimensions is rated on a 4-point scale, where 0 for having more than three clinically significant errors or omissions, 1 for having three clinically significant errors or omissions, 2 for having two clinically significant errors or omissions, 3 for having one clinically significant error or omission, and 4 for having no clinically significant errors or omissions. As shown in Fig.~\ref{fig:human_eval_report}, 
the reports from our \OURMODEL~demonstrate the fewest or no clinical errors or omissions.

\noindent \textbf{Grounded Report Generation.} 
On grounded report generation~(the `\OURDATASETWOS-Region' part of Tab.~\ref{tab:rg_statistic}), we decompose the original report into specific brain anatomical regions with more details in Sec.~\ref{sec:report_generation_module_training}, 
serving as the ground truth during this evaluation. 
\OURMODEL~pre-trained on \RELEASERG~and \OURDATASET~outperforms the baselines significantly. It achieves a RadGraph score of 32.20$\pm$24.18\% , a RadCliQ score of 0.32$\pm$0.65, and a RaTEScore of 59.53$\pm$21.12\%. 
It is important to note that the grounded report generation performance of the baselines is inferior to their global report generation performance, suggesting that existing models cannot describe regions accurately.

\begin{figure}[!htb]
\centering
\includegraphics[width=\textwidth]{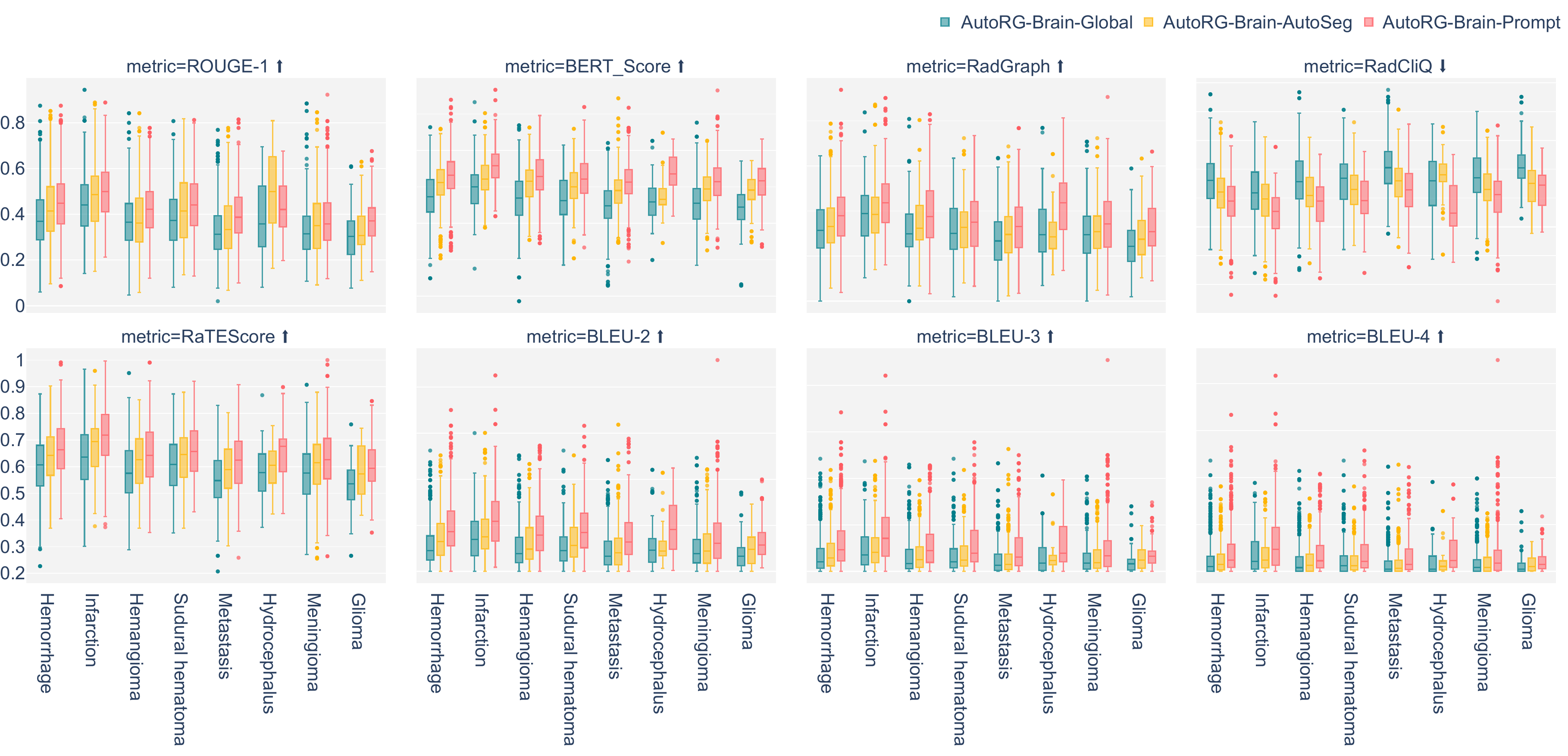}
\caption{The comparison results between the ground truth global report and global reports written by \OURMODELWOS-Global, \OURMODELWOS-AutoSeg, and \OURMODELWOS-Prompt on \OURDATASET~dataset. ROUGE-1, Bert-Score, RadGraph, RadCliQ, RaTEScore, BLEU-2, BLEU-3, and BLEU-4 are reported.}
\label{fig:report_quality}
\end{figure}

\noindent \textbf{Ablation Studies.}
The ablation results of three inference strategies have been showcased in Tab.~\ref{tab:rg_statistic}, the following phenomenon can be observed:
(i) \OURMODELWOS-Prompt and \OURMODELWOS-AutoSeg demonstrate better performance than \OURMODELWOS-Global as shown in Fig.~\ref{fig:report_quality}, which suggests the efficiency of integrating grounded visual clues into global report generation. 
(ii) With the human prompting, {\em i.e.}, simply specify the regions of interest on the structure segmentation results,
\OURMODELWOS-Prompt reaches the best performance. 
It indicates the refinement of grounding clues can improve the overall report quality. 
In terms of the impact of \OURDATASET~and \RELEASERG~as pre-training datasets, we can also draw the following observations: 
(i) Incorporating \RELEASERG~in pre-training enhances global and regional performance in report generation for \OURDATASET~dataset, 
likely due to the high-quality segmentation-report pairs in the \RELEASERG~dataset. (ii) However, incorporating the \OURDATASET~in pre-training does not benefit the evaluation on \RELEASERG. We conjecture this is because \OURDATASET, collected from one medical center, 
dominates around 90\% of the training data, resulting in potential bias in the dataset, thus showing slightly inferior performance on a diverse and multi-center test set.

\begin{figure}[!htb]
\centering
\includegraphics[width=\textwidth]{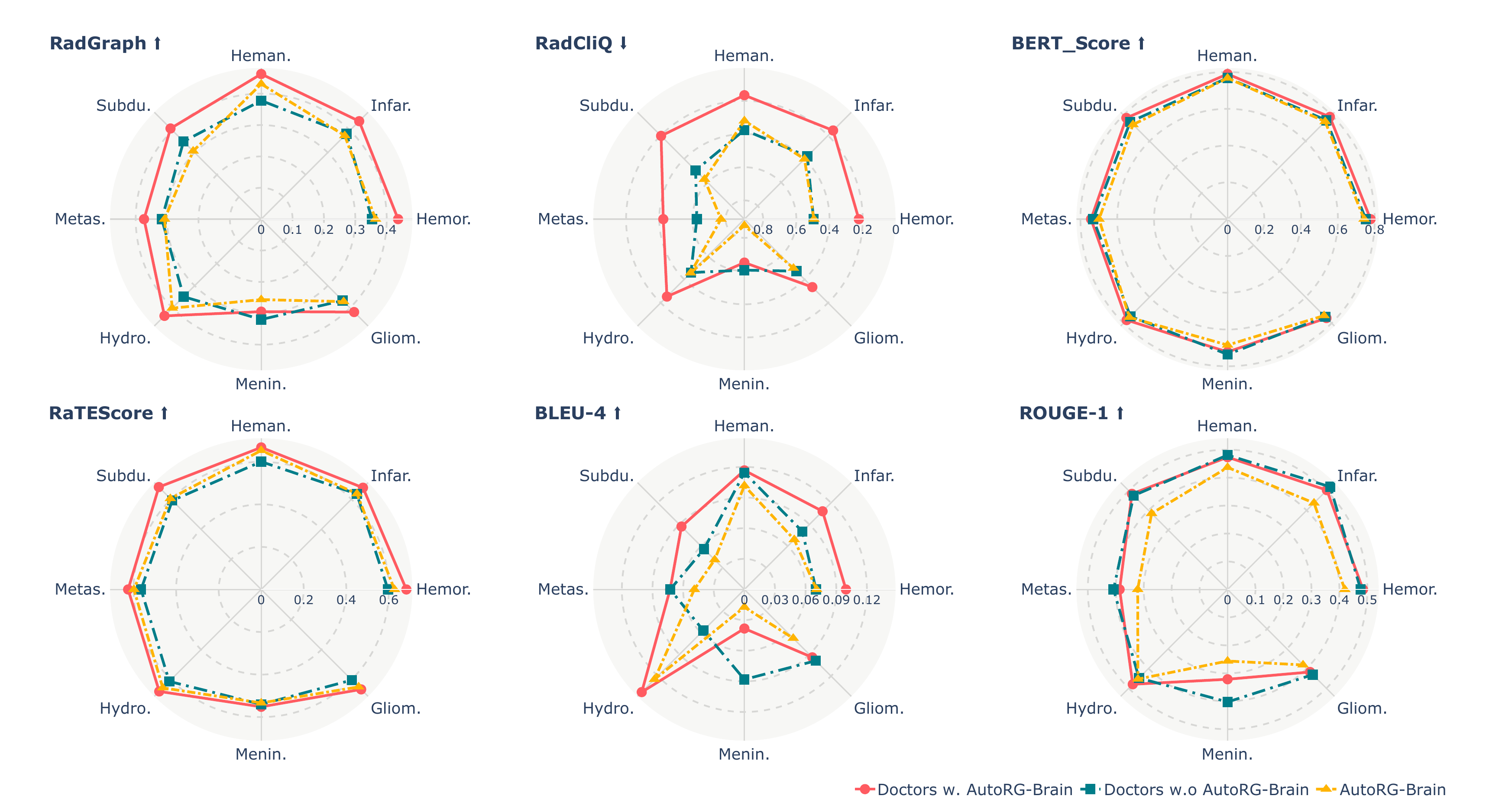}
\caption{The comparison results between the ground truth report and reports written by \OURMODELWOS, doctors w.o \OURMODELWOS, and doctors w. \OURMODEL~detailed on various diseases (Heman.: Hemangioma, Infar.: Infarction, Hemor.: Hemorrhage, Gliom.: Glioma, Menin.: Meningioma, Hydro.: Hydrocephalus, Metas.: Metastasis, Subdu: Sudural hematoma). RadGraph, RadCliQ, Bert-Score, RaTEScore, BLEU-4, and ROUGE-1 are reported.}
\label{fig:clinical_trail}
\end{figure}

\subsection{Evaluation on Real Clinical Scenario}
\label{sec2_3}
To evaluate the ability of \OURMODEL~in real clinical scenarios, we integrate our system into the radiology daily routines to assess its usefulness in the report writing process. We select 200 cases from \OURDATASET~with ground truth reports written by senior radiology experts with 6-8 years of experiences. Two groups of radiologists are involved in this experiments. Group A is set to mimic the real clinical setting, 
where \textbf{three junior radiologists (2-5 years)} are shown Brain MRI of T1WI, T2WI, T2FLAIR, and DWI modalities and a standard report template used in daily diagnosis. 
Group B is the model-integrated setting, where \textbf{another three junior radiologists (2-5 years)} are shown Brain MRI of the same four modalities as group A, the corresponding anomaly segmentation masks generated by our segmentation tool, and the report generated by \OURMODELWOS. Both groups of radiologists are asked to edit the given template based on the image scans.
The annotation page is shown in Appendix Fig.~\ref{fig:report_writing_annotation}. 
The comparison results between the ground truth report and those written by doctors in group A and group B are presented in Appendix Tab.~\ref{tab:human_accuracy} with details on various diseases shown in Fig.~\ref{fig:clinical_trail}. It shows that the integration of \OURMODEL~leads to significant improvements in report writing quality, especially on the most relevant metric to radiology report generation, such as RadGraph, RadCliQ, and RaTEScore. 
\section{Discussion}\label{sec:discussion}

In this paper, we propose \OURMODELWOS, the first brain MRI interpretation system for grounded report generation, that can be prompted by segmentation of abnormality or structures. The primary goal is to enhance the efficiency of radiologists' daily workflows in real clinical settings.

\noindent \textbf{The First Regional Brain MRI Interpretation System.} 
Our proposed \OURMODEL~integrates diverse datasets to train a comprehensive brain MRI interpretation system. The system provides a detailed report for each auto-segmented or human-prompted abnormal region along with the pixel-wise grounding visualization. Results show that our system achieves state-of-the-art performance on brain anomaly segmentation, brain structure segmentation, global report generation, and grounded report generation. When applied to real clinical scenarios, our system shows the potential to enhance radiologists' efficiency. With the fast development of generalist medical AI (GMAI)~\cite{moor2023foundation}, our system can serve as an agent to fill the gap of brain MRI interpretation within the GMAI framework. 

\noindent \textbf{The Two-stage Decomposition is Critical.} 
The key point of integrating grounding into global report generation is to decompose the report generation task into two steps, 
\emph{i.e.}, first localizing anomaly regions, 
then generating regional reports, 
and consolidating them into a comprehensive global report. 
Results in Tab.~\ref{tab:rg_statistic} show its superior performance than one-step global-feature-based report generation. 
This is likely because of three reasons:
(i) Report generation based on the global feature fails to capture fine-grained information which is crucial in improving accuracy. As a consequence, it may lead to a mismatch of lesion location and report content during testing, especially in small-lesion, rare lesion location situations; 
(ii) Segmentation masks provide attention on the global image feature, which reduces interference from unimportant areas, allowing report generation with more precise information; 
(iii) The report generated based on global features remains fixed once the model is trained. In contrast, the report generated using regional clues can be enhanced through human prompting or improvements in the segmentation model. Cases of generated reports by \OURMODELWOS-Global, \OURMODELWOS-AutoSeg, and \OURMODELWOS-Prompt are shown in Appendix Sec.\ref{sec:RG_three_setting}.

\noindent \textbf{Training on Large-scale Partially Labeled Data is Effective.} 
Due to the scarcity of large-scale datasets with complete labels, 
we show that training the model can leverage diverse datasets with partial annotations, including public or private hospital datasets, by compensating them with synthetic supervision. 
As a result, our model enables to: (i) Identify anomalies across different brain MRI datasets, regardless of specific pathology or condition; 
(ii) Identify brain structures on multi-modal brain MRIs with or without anomalies. Additionally, the inclusion of large-scale image-report pairs from hospitals greatly aids in report-generation training. 
We leverage our trained segmentation module and GPT-4 to process the image-report pair at global and regional levels to facilitate report generation on diverse visual prompts. 

\noindent \textbf{\OURMODEL~Helps Radiologists in Real Practice.}
In clinical impacts, \OURMODEL~has shown promising help for radiologists by integrating into real clinical workflows. AI-powered junior radiologists produce reports that are closer to senior radiologists. 
This improvement is attributed to two factors: (i) Reduction of false negatives through automatic abnormal region segmentation, preventing critical findings from being missed. (ii) Comprehensive reports with detailed information, including characteristics such as signal, shape, location, deformation, displacement, and other relevant features of identified regions. Appendix Sec.~\ref{sec:case_ai_assisted} provides specific case illustrations. Our experiments highlight the potential benefits of AI integration into radiologists' workflow, which enables grounded rationales in the generated reports, and fosters collaboration between humans and AI, providing valuable insights into building trustworthy AI systems that genuinely assist doctors.

\noindent \textbf{A New Dataset and Benchmark for Grounded Report Generation.}
We propose to add new annotations to the publicly available anomaly segmentation datasets, termed as \textbf{\RELEASERGWOS}. 
To our knowledge, this is the first open-source collection to provide paired images with detailed reports and anomaly segmentations. 
In the experiment, \OURMODEL~showcases the effectiveness of segmentation as a preliminary step to provide grounded visual clues for image interpretation.  We hope the open-source \RELEASERG~dataset will facilitate on grounded image interpretation research such as grounded report generation or grounded diagnosis and accelerate the progress and development of the medical imaging community.

\noindent \textbf{Limitations.}
While our work primarily focuses on constructing an effective image interpretation system for brain MRI which benefits clinicians' workflow, 
we encountered certain limitations:
{\em Firstly}, the quality of segmentation masks can impact report quality. 
To address this, we plan to incorporate more anomaly segmentation datasets to train segmentation module and refine interface for realtime radiologist interaction;
{\em Secondly}, our current approach relies solely on images from four modalities, other valuable information, such as patient history or data from alternative modalities.  Moving forward, we can explore incorporating additional input information to potentially enhance the overall system performance;
{\em Lastly}, our current system currently supports report generation, 
in future work, we envision leveraging multi-modal data for more accurate and comprehensive diagnoses, treatment planning.

\section{Methodology}
\label{sec:method}

\subsection{Algorithm}
\begin{figure}[!ht]
\centering
\includegraphics[width=\textwidth]{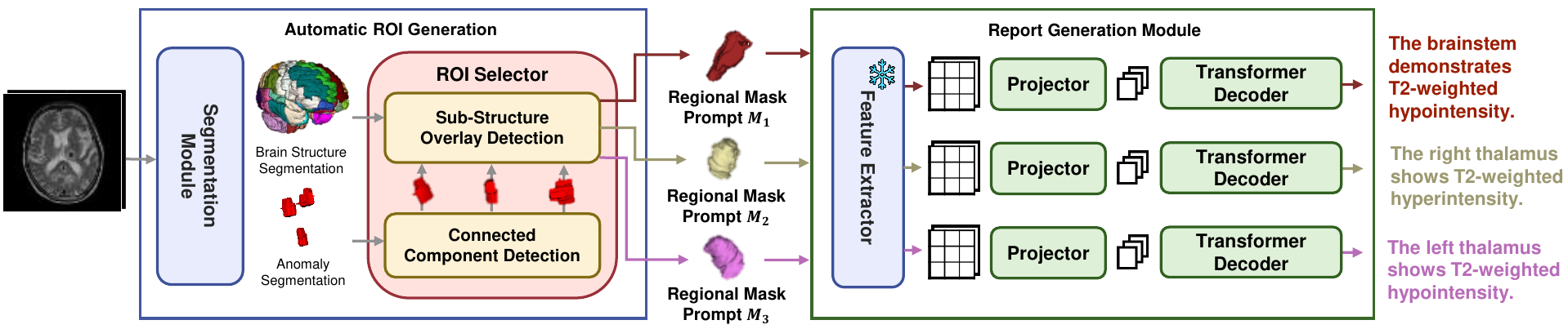}
\caption{The architecture of our proposed \OURMODELWOS. The segmentation module ($\Phi_\text{SEG}(\cdot)$) produces global brain structure and anomaly segmentations. Subsequently, for each connected component of the anomaly segmentation, the ROI selector generates overlaid sub-structures as regional prompts. The report generation module ($\Phi_\text{RG}(\cdot)$) extracts features for each provided regional prompt. The regional features are then projected into input tokens, and the transformer decoder-based module generates the corresponding findings. The combination of these local findings forms the final global finding.}
\label{fig:sub_module_arch}
\end{figure}

Our proposed system aims to automate the interpretation of brain MRI images, by identifying anomalies, pinpointing the affected brain regions, and generating the corresponding reports as shown in Fig.~\ref{fig:sub_module_arch}. 
\OURMODEL~comprises two components: 
(i) an automatic ROI generation component including a trainable segmentation module~($\Phi_\text{SEG}(\cdot)$) and a rule-based ROI selector, and (ii) a visual prompting guided report generation component~($\Phi_\text{RG}(\cdot)$). 
The region of interest generation stage uses the segmentation module to densely segment the anomalies and brain structures,
and an ROI selector to detect abnormal regions. The report generator will then describe the abnormal regions based on mask prompts, leveraging global and regional features. In the following sections, we will describe the development of the two components in detail.

\subsubsection{Segmentation Module Training}\label{sec:segmentation_training}
\begin{figure}[!ht]
\centering
\includegraphics[width=\textwidth]{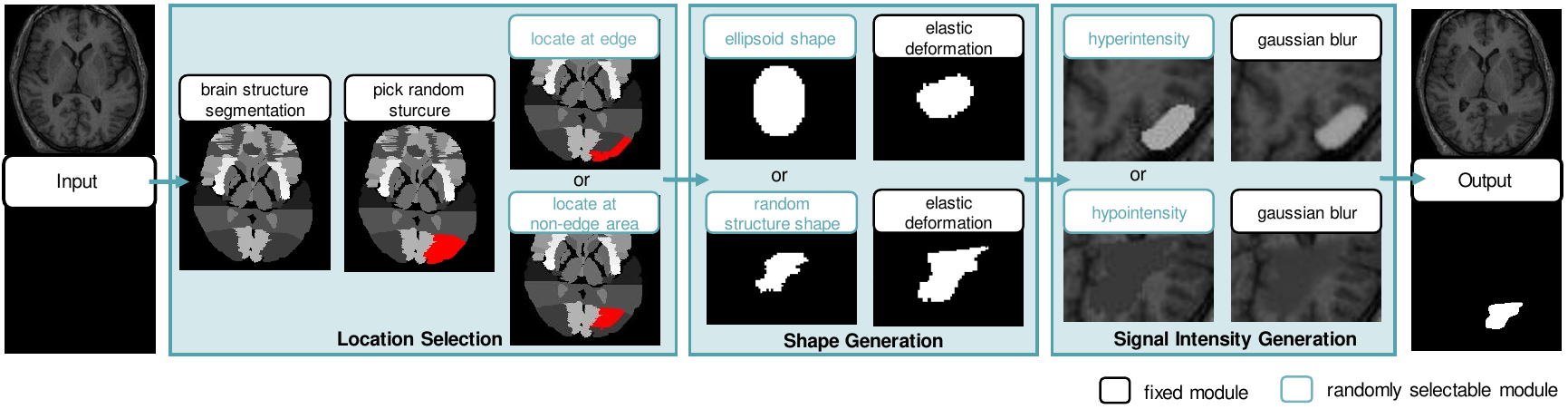}
\caption{Synthetic anomaly generation. Firstly, we randomly select a sub-structure's edge or non-edge area as the location. Secondly, we initiate the anomaly shape from either ellipsoid or a random brain structure's shape, and perform elastic deformation on it to generate the final shape. Lastly, we randomly generate the hyperintensity or hypointensity texture and adopt Gaussian Blur to fuse the `anomaly' into the original image.}
\label{fig:synthetic_anomalies}
\end{figure}

The segmentation module aims to perform segmentation for both anomaly and brain structures for the scan in any of the four common brain MRI modalities: T1-weighted, T2-weighted, FLAIR, and DWI.
In this paper, we adopt a two-stage training pipeline using a large-scale in-house dataset from hospitals, and existing public segmentation datasets (datasets details shown in Sec~\ref{sec:dataset_overview}): (i) self-supervised training with healthy samples from in-house hospital data and (ii) semi-supervised training with both hospital healthy samples and public datasets. 
The detailed process is as follows:

\noindent \textbf{Self-Supervised Training Stage.} 
\label{sec:self-sup-train-stage}
At this stage, our goal is to train the segmentation module from scratch using a large-scale multi-modal hospital dataset without any segmentation labels. To get the brain structure supervision, we register the publicly available MRI atlas with brain structure labels~\cite{faillenot2017macroanatomy} to the unlabeled hospital's healthy brain MRIs using the EasyReg algorithm~\cite{iglesias2023ready}, the transformed structure masks are treated as ground-truth for training  structure segmentation branch. On the other hand, for anomaly segmentation, we introduce an approach that generates synthetic anomalies based on report descriptions and blends them into the healthy brain scans, resulting in paired abnormal image-mask training samples. 
We simulate lesions by altering signal intensity in specific brain structures, for example, a report may describe anomalies as ``patchy abnormal signals in the left basal ganglia, with hypointensity on T1W images and hyperintensity on T2W, FLAIR, and DWI images.'' 
In detail, our signal-aware anomaly modeling involves three steps: (i) location selection, (ii) shape generation, and (iii) signal intensity generation. The procedure is illustrated in Fig.~\ref{fig:synthetic_anomalies}.

\vspace{-0.2cm}
\begin{itemize}
\setlength\itemsep{0.15cm}

    \item  \textbf{Location Selection.} 
    We first decide where to put the synthetic anomaly in a certain MRI scan. Since, in clinical practice, report findings often employ brain structure to describe the locations of anomalies, we simply pick a random brain structure as the location of the `lesion'. Considering some lesions may be at the margin of brain structures, such as subdural hematoma or meningioma, we will perform Morphological Gradient~\cite{Poorter1999GrowthRO} calculation to detect the edge of the brain structure, pick a random point either from the edge or non-edge area within the brain structure as the center point of the `lesion'.
    
    \item  \textbf{Shape Generation.} 
    We then initiate the general shape of the anomaly.
    In existing works~\cite{zhang2023self,hu2023label}, 3D ellipsoids with arbitrary long and short axes are commonly used, then followed by elastic transformations to generate lesion shapes. 
    However, this method was primarily designed for tumor lesions. 
    To simulate a broader range of lesion shapes, we also incorporate various 3D brain anatomical regions from our registered mask as initial shapes, which then undergo elastic transformations to create diverse shapes.
   
    \item  \textbf{Signal Intensity Generation.} 
    We inpaint the anomaly by simulating the abnormal intensity distribution. An abnormal signal typically indicates a relatively high, relatively low, or mixed signal intensity compared to the intensity of the original healthy brain area. 
    Based on such prior, we first calculate some reference intensity values for the original normal region. Assuming we are going to put the anomaly in the region $R$, then we can naturally obtain:
    \begin{equation}
    \begin{aligned}
        I_\text{avg}=\text{AVG}\{I_{(x, y, z)} | (x, y, z)\in R\}, \\
    \end{aligned}
    \end{equation}
    where $I_{(x,y,z)}$ denotes the voxel intensity at index $(x,y,z)$, $\text{AVG}(\cdot)$ denotes the average intensity of the selected voxels. Similarly, we can obtain $I_\text{max}$ and $I_\text{min}$ by adopting maximize or minimize function to all voxels. 
    Based on the fundamental intensity statistics indicators, 
    we can define the anomaly intensity distributions. 
    If it is hyperintensity, then we will randomly pick an intensity value from a uniform distribution:
    \begin{equation}
        I_a \sim \mathcal{U}(I_\text{avg}+\varepsilon, I_\text{max}),
    \end{equation}
     where $I_a$ denotes the anomaly intensity value at the center point, $\mathcal{U}(\cdot)$ denotes the uniform distribution and $\varepsilon$ is a hyperparameter to control the anomaly intensity shifting. On the other hand, if the anomaly intensity is hypointensity, the intensity value will be picked as:
    \begin{equation}
        I_a \sim \mathcal{U}(I_\text{min}, I_\text{avg}-\varepsilon).
    \end{equation}
    Lastly, we employ Gaussian Blur with $\sigma_{b}$ as the standard deviation, $I_a$ as the highest voxel intensity, and voxel-wise Euclidean distance as the distance measurement to fill in the whole anomaly regions and replace the original corresponding normal areas with it.
\end{itemize}

In practice, it is common for patients to have multiple lesions simultaneously. Therefore, we generate a random number of lesions within the brain MRI, each following the aforementioned steps.  
The above self-supervised approach enables the model to learn to identify various anomalies and brain structures in multi-modal healthy and abnormal brain MRIs.

\noindent \textbf{Semi-Supervised Training Stage.} 
At the second training stage, we further integrate publicly available brain anomaly segmentation datasets, for further refinement. Specifically, we use our pre-trained model to generate brain structure labels for the public datasets, and use the provided lesion annotations as anomaly ground truth.

\noindent \textbf{The Network Backbone and Loss Function.}
We employ the nnU-Net framework~\cite{isensee2021nnu} as our core segmentation architecture. Specifically, our model utilizes distinct encoders for various imaging modalities, and share the same decoder. 
Each input image is processed by the modality-specific encoder, and the shared decoder subsequently generates outputs for both anomaly and brain structure segmentations. The loss function of the segmentation task is as follows:
    \begin{equation}
    \begin{aligned}
    \mathcal{L}_{\text{seg}}(s,\hat{s})=\lambda_{1}\cdot(1-\frac{|s|+|\hat{s}|}{2\cdot |s\bigcap \hat{s}|})-\lambda_{2}\cdot\sum s\log \hat{s}, \\
    \end{aligned}
    \label{seg_loss}
    \end{equation}
where $s$ is the label for supervision and $\hat{s}$ is the predicted binary mask. $\lambda_{1}$ is the factor for dice loss while $\lambda_{2}$ is the factor for cross entropy loss. The total loss for the segmentation module is as follows:
    \begin{equation}
    \begin{aligned}
    \mathcal{L}_{\text{SEG}} = \mathcal{L}_{\text{seg}} (s_{a},\hat{s}_{a})+\mathcal{L}_{\text{seg}} (s_{b},\hat{s}_{b}),
    \end{aligned}
    \label{seg_loss}
    \end{equation}

where $s_{a},\hat{s_{a}}$ represent the anomaly segmentation label and prediction, while $s_{b},\hat{s_b}$ represent the brain structure segmentation label and prediction. Notably, $s_b$ includes multiple classes denoting different brain structures and we adopt class-wise averaging to get the brain structure segmentation loss $\mathcal{L}_{\text{seg}} (s_{b},\hat{s}_{b})$.

\subsubsection{Report Generation Module Training}\label{sec:report_generation_module_training}
After training the segmentation module, we freeze it and train the report generation module. The ground truth supervision includes both global and regional reports as introduced in Sec.~\ref{sec:implement_details}. The model is trained to output multi-granular reports according to different visual mask prompts from ROI selector.  

\noindent \textbf{Training Stage.}
An ideal training sample would require $(X, M, T)$, 
denoting the original MRI scan, the regional mask prompt, 
and the ground truth report describing the region. 
Our curated \RELEASERG~can satisfy such condition, with the anomaly annotation mask as $M$ and the corresponding radiologist-written report as $T$. While the other in-house dataset \OURDATASETWOS, only has $(X, T_{g})$ pairs, where $T_{g}$ is the global report. We first compensate for these by leveraging the trained segmentation tool and GPT-4~\cite{achiam2023gpt}. In detail, we decompose $T_{g}$ into modal-wise regional descriptions $T_{r}$ for different brain structures by GPT-4 using prompt in Appendix Sec.~\ref{prompt_report_decomp}, a processed region-report pair example is as follows: 

\begin{mdframed}[backgroundcolor=gray!10] 
``Bilateral Frontal Parietal Lobes'': \{ \\ 
    ``FLAIR'':\ ``Multiple punctate lesions with hyperintensity are visible in the bilateral frontal parietal lobes'', \\
    ``T1'':\ ``Multiple punctate lesions with isointensity are visible in the bilateral frontal parietal lobes'',\\
    ``T2W'':\ ``Multiple punctate lesions with hyperintensity are visible in the bilateral frontal parietal lobes'' 
    \}.
\end{mdframed}

Then, the pre-trained segmentation module is employed to generate an anomaly mask and a brain structure segmentation mask. Based on the two masks and $T_r$, we can simulate and compensate a series of regional mask prompts $M$ for the training stage. 
Specifically, for a certain $T_r$, we make up $M$ with the predicted masks of brain structures mentioned in $T_r$. For example, for the above   ``Bilateral Frontal Parietal Lobes'' $T_r$ case, $M$ will be synthesized as the union of the dense segmentation masks on the left frontal lobe, right frontal lobe, left parietal lobe, and right parietal lobe brain structures.
\textbf{Notably}, we also maintain a special ``regional'' case where $M$ is an all-one mask and the corresponding report is just global report $T_g$, indicating a special prompting case where the whole ``region'' should be described. For simplicity, next, we will not distinguish whether a supervision report is regional or global, typical training cases are uniformly denoted as $(X, M, T)$ where $T$ can be of arbitrary granularity as long as it corresponds to the inputs visual prompt $M$.

\noindent \textbf{The Network Backbone and Loss Function.}
Our report generation module mainly contains three sub-modules, \emph{i.e.}, an encoding module $\Phi_\text{Enc}$, a projection module $\Phi_\text{Proj}$, and a report decoding module $\Phi_\text{Dec}$. 
We re-use the whole encoder and the first three decoder layers in the frozen segmentation module as the encoding module to embed the input scan $X$:
    \begin{equation}
    f_{g}=\Phi_\text{Enc}(X)  \in \mathbb{R}^{h \times w \times d \times dim},
    \end{equation}
where $f_{g}$ is the global feature map for the whole scan and $h, w, d$ is the map size, $dim$ denotes the embedding dimension. 
We further obtain the regional features by element-wise product between the global feature and the binary mask prompt:
\begin{equation}
    f_{l} = f_{g} \odot M \in \mathbb{R}^{h \times w \times d \times dim},
\end{equation}
where $\odot$ is the element-wise production function and $M$ is the region of interests mask. Then, we concatenate the global and regional features on the dimension-wise, resulting in $f_{con} \in \mathbb{R}^{h \times w \times d \times 2dim}$. A CNN-based projection module is adopted here, to further mix and compress the feature embedding and output a sequence of visual tokens by simply flattening:
\begin{equation}
    f_{m} = \Phi_\text{Proj}(f_{con}) \in \mathbb{R}^{P \times dim},
\end{equation}
where $f_{m}$ is the final mixing features, $P = h' \cdot w'  \cdot d'$ is the patch length, and $h', w', d'$ are the original feature map size before flattening. Finally, we decode out the report based on the mixing features.
Our report generation module is a typically decoder-only transformer based on self-attention as GPT-series~\cite{radford2019language}, where we view $f_{m}$ as sequential input and decode out the expected word tokens. We adopt the classical auto-regressive loss as the final optimization objective:
\begin{equation}
    \mathcal{L}_{\text{RG}} = - \sum_i \log(\Phi_\text{RG}(T_i)|T_{<i},x),
\end{equation}
where $T_i$ is the $i$-th token in groudtruth report and $T_{<i}$ denotes all the word token before it. 

\subsubsection{Inference Time}\label{sec:inference}
At inference time, our model can be used under two modes, 
\emph{i.e.}, human interaction mode and automatic mode.
In the former, clinicians can interact with our system with any desired ROI masks, or pick from the segmentation masks to directly generate reports, termed as \OURMODELWOS-Prompt, which can hardly be realized by most former report generation works. 
In the latter, the model receives a 3D MRI image as input and the segmentation module will generate a series of anomaly region prompts. 
Then the report generation will predict regional reports one by one. Concatenating them together and filling the normal parts with standard templates, we can get a global report with linked regional visual clues for each subparagraph, termed as \OURMODELWOS-AutoSeg. It is worth emphasizing that, we can also predict the global report without any decomposition as former classical report generation works, just by inputting an all-one mask prompt to our report generation module, termed as \OURMODELWOS-Global. However, compared with our proposed default regional-generation way, this is much less interpretable and also poorer in final performance as shown by our results in Sec.~\ref{sec2_2}.

\subsubsection{Implementation Details}\label{sec:implement_details}
In this section, we delve into the implementation details in the segmentation and report generation training process. 
We conduct all our experiments using \texttt{PyTorch} framework and \texttt{Transformers} python package.  

\textbf{Acquiring Segmentation Labels.} 
For the healthy samples from \OURDATASET~used in both stages, 
the brain structure label is given by the SOTA registration method EasyReg~\cite{iglesias2023ready} integrated in FreeSurfer 7.4 with `a01' sample from Hammers-n30r95 as source. The anomaly segmentation label is generated on the fly using methods in Sec.~\ref{sec:self-sup-train-stage}. For the public datasets used in the second stage, the brain structure label is given by the first-stage model and the anomaly segmentation label is the original annotation in those datasets.

\textbf{Segmentation Training}. 
The encoder and decoder backbone, data pre-process, and data augmentation procedure follow the nnU-Net\footnote{https://github.com/MIC-DKFZ/nnUNet/tree/nnunetv1}~\cite{isensee2021nnu} framework. During training, we establish a global batch size of 2 and input patch size of $(112, 128, 112)$. We use the poly learning rate scheduler with initial learning rate as 1e-2 and the exponent as 0.9. We use the \texttt{torch.optim.SGD} optimizer with the weight decay as 3e-5 and momentum as 0.95. We train the segmentation model for a maximum of 1k epochs.  

\textbf{Acquiring Report Generation Labels}: 
When processing the report labels of our \OURDATASET~dataset, 
for each 3D image of a specific modality, we acquire two types of reports: global reports and regional reports. We first get a model-wise regional report for different groups of brain structures using the prompt in Appendix Sec.~\ref{prompt_report_decomp}. Then we concatenate all the regional reports on a specific modality to form the global report of that modality. For our \RELEASERG~dataset, the radiologists write a report for each patient case, then we use the prompt shown in Appendix Sec.~\ref{prompt_report_decomp} to decompose the report into a modal-wise report using GPT-4. We use the decomposed modal-wise report as our ground-truth global report for each image.

\textbf{Report Generation Training}. 
During further auto-regressive training on \OURDATASETWOS, 
the optimization objective aligns with that of language generation task. The report generation backbone is the 355M-parameter model GPT-2 Medium~\cite{radford2019language} fine-tuned on PubMed abstracts~\cite{papanikolaou2020dare}. 
During training, we set the global batch size of 4 and maintain a maximum context length of 1024 tokens. 
We use \texttt{torch.optim.lr\_scheduler.ReduceLROnPlateau} as the learning rate scheduler with the initial learning rate as 5e-5 and \texttt{torch.optim.AdamW} as the optimizer. The model is trained for 1k epochs on one A100 GPU, which takes about 5 days.

\subsection{Dataset}\label{sec:dataset_overview}
As it is uncommon for a single dataset to contain all the necessary information for training our model, {\em i.e.}, images, reports, and anomaly and anatomical segmentation masks, 
we adopt multiple datasets in the training and testing process, 
with each of the dataset reflects part of the functions in our proposed model. In the following, we give detailed descriptions on each dataset and summarise the usage of all datasets in Tab.~\ref{tab:dataset_summary}.

\begin{table}[!t]
\centering
\caption{Overview of the datasets used in the training and testing stage of \OURMODEL. \xmark~means no label, \cmark~means having the ground-truth label, \cmark$^{*}$~means having the pseudo label.}
\vspace{3pt}
\renewcommand{\arraystretch}{1.1}
\footnotesize
\resizebox{1.0\linewidth}{!}{
\begin{tabular}{c|c|c|ccc|ccc}
\toprule
\multirow{2}{*}{Module} & \multirow{2}{*}{Dataset} & \multirow{2}{*}{Modality} &\multicolumn{3}{c|}{Annotation Type} &\multicolumn{3}{c}{Image Split} \\
\cmidrule(lr){4-6} \cmidrule(lr){7-9}
& & & \multicolumn{1}{c|}{Reports} & \multicolumn{1}{c|}{\makecell{Anomaly \\ Seg.}} & \makecell{Anatomical \\ Seg.} & \multicolumn{1}{c|}{Training}  & \multicolumn{1}{c|}{Validation} & Test \\
\midrule
\multirow{3}{*}{\makecell{\\ Segmentation \\ - First Stage}} & \makecell{\OURDATASET~(Ours) \\ - Healthy Data} & \makecell{T1WI, T2WI\\T2FLAIR, DWI} & \cmark & \cmark$^{*}$ & \cmark$^{*}$ & 8,657 & 2,167 & - \\
\cmidrule(lr){2-9}
 & \makecell{ISLES2022} & \makecell{DWI} & \xmark & \cmark & \xmark & - & - & 247 \\
 \cmidrule(lr){2-9}
 & \makecell{BraTS2021} & \makecell{T2FLAIR} & \xmark & \cmark & \xmark & - & - & 1,251 \\
\midrule
\multirow{11}{*}{\makecell{\\ \\ \\ \\ \\ \\ Segmentation \\ - Second Stage}} & \makecell{\OURDATASET~(Ours) \\ - Healthy Data} & \multirow{3}{*}{\makecell{\\ \\ T1WI, T2WI\\T2FLAIR, DWI}} & \cmark & \cmark$^{*}$ & \cmark$^{*}$ & 8,657 & 2,167 & - \\
  \cmidrule(lr){2-2}   \cmidrule(lr){4-9}
 & \makecell{\OURDATASET~(Ours)\\- Abnormal Data} &  & \xmark & \xmark & \xmark & - & - & 100 \\
  \cmidrule(lr){2-2}   \cmidrule(lr){4-9}
 & RP3D-Brain &  & \xmark & \xmark & \xmark & - & - & 100 \\
\cmidrule(lr){2-9}
& \makecell{ISLES2022} & \makecell{DWI} & \xmark & \cmark & \cmark$^{*}$ & 200 & 50 & - \\
\cmidrule(lr){2-9}
& \makecell{BraTS2021} & \multirow{5}{*}{\makecell{\\ \\ T1WI, T2WI \\ T2FLAIR}} & \xmark & \cmark & \cmark$^{*}$ & 3000 & 753 & - \\
  \cmidrule(lr){2-2}   \cmidrule(lr){4-9}
& \makecell{BraTS-MEN} & & \xmark & \cmark & \cmark$^{*}$ & 2400 & 600 & - \\
  \cmidrule(lr){2-2}   \cmidrule(lr){4-9}
& \makecell{BraTS-MET} & & \xmark & \cmark & \cmark$^{*}$ & 570 & 144 & - \\
  \cmidrule(lr){2-2}   \cmidrule(lr){4-9}
& \makecell{BraTS-PED} & & \xmark & \cmark & \cmark$^{*}$ & 237 & 60 & - \\
  \cmidrule(lr){2-2}   \cmidrule(lr){4-9}
& \makecell{BraTS-SSA} & & \xmark & \cmark & \cmark$^{*}$ & 144 & 36 & - \\
\cmidrule(lr){2-9}
& \makecell{ATLAS} & \multirow{2}{*}{T1WI} & \xmark & \cmark & \cmark$^{*}$ & 524 & 131 & - \\
\cmidrule(lr){2-2} \cmidrule(lr){4-9}
& \makecell{Hammers-n30r95} & & \xmark & \cmark$^{*}$ & \cmark & 20 & 4 & 6 \\
\cmidrule(lr){2-9}
& \makecell{WMH} & \makecell{T1WI \\ T2FLAIR}& \xmark & \cmark & \cmark$^{*}$ & 96 & 24 & - \\
\midrule
\multirow{2}{*}{\makecell{\\ Report Generation \\ Module}} & \OURDATASET~(Ours) & \makecell{T1WI, T2WI \\ T2FLAIR, DWI} & \cmark & \cmark$^{*}$ & \cmark$^{*}$ & 24,613 & 2,155 & 4,025 \\
\cmidrule(lr){2-9}
& \RELEASERG~(Ours) & \makecell{T1WI, T1CE\\T2WI, T2FLAIR\\DWI, ADC} & \cmark & \cmark & \cmark$^{*}$ & 2,382 & 338 & 688 \\
\bottomrule
\end{tabular}}
\label{tab:dataset_summary}
\end{table}

\subsubsection*{Report Generation}

\vspace{-0.2cm}
\begin{itemize}
\setlength\itemsep{0.5em}
\item[$\bullet$] \textbf{Shanghai Six People's Hospital~(SSPH)}: 
We collected 12,535 cases of 30,793 brain MRI imaging-report pairs at Shanghai Sixth People's Hospital~(SSPH)\footnote{This study has been approved by the Ethics Committee of Shanghai Sixth People's Hospital [IRB code: 2023-KY-082 (K)].}. All MRI scans were performed on one of the eight MRI scanners (GE Medical Systems Signa Pioneer, Philips Medical Systems EWS, Philips Medical Systems Ingenia, Siemens Prisma, Siemens Skyra, Siemens Verio, UIH uMR 780, UIH uMR 790) from January 2019 to March 2023 at Shanghai Sixth People's Hospital. Each data sample consists of one case report and scans of four axial MRI modalities, namely, T1WI, T2WI, T2FLAIR, and DWI. We use 2,707 healthy cases of 10,824 images to pre-train our segmentation model in a self-supervised manner, with 2,165 cases of 8,657 images as trainset and 542 cases of 2,167 images as testset. 
We use all 12,535 cases of 30,793 images to train our report generation model, with 10,028 cases of 24,613 images as trainset, 539 cases of 2,155 images as valset, and 1,968 cases of 4,025 images as testset. 

\item[$\bullet$] \textbf{\RELEASERGWOS}: The dataset comprises 1,007 cases of 3,408 imaging-report pairs. The image data are sourced from five well-known public anomaly segmentation datasets, including the ISLES2022, White Matter Hyperintensity Challenge Dataset (WMH), BraTS2021, BraTS-MEN, and BraTS-MET, covering 6 MRI modalities (T1-weighted, T2-weighted, DWI, T2-Flair, ADC, and T1-contrast) and 5 distinct disease types (infarction, white matter hyperintensity, glioma, meningioma, and metastasis). We invite five radiologists (4-5 years) to write report findings and impressions for the annotated anomaly area of each patient case. The training, validation, and testing split along with the annotation types are shown in Tab.~\ref{tab:brainrg_summary}. 

\begin{table}[!htb]
\centering
\caption{The annotation type and image-level training, validation, and testing split of the \RELEASERG~dataset. \cmark~means having the ground truth label, \cmark$^{*}$~means having the pseudo label.}
\vspace{3pt}
\renewcommand{\arraystretch}{1.1}
\setlength{\tabcolsep}{4pt}
\footnotesize
\begin{tabular}{c|c|c|ccc|ccc}
\toprule
\multirow{2}{*}{Dataset} & \multirow{2}{*}{Sub-Dataset} & \multirow{2}{*}{Modality} &\multicolumn{3}{c|}{Annotation Type} &\multicolumn{3}{c}{Image Split} \\
\cmidrule(lr){4-6} \cmidrule(lr){7-9}
& & & \multicolumn{1}{c|}{Reports} & \multicolumn{1}{c|}{\makecell{Anomaly \\ Seg.}} & \makecell{Anatomical \\ Seg.} & \multicolumn{1}{c|}{Training}  & \multicolumn{1}{c|}{Validation} & Test \\
\midrule
\multirow{5}{*}{\makecell{\\ \\ \RELEASERG}} & \makecell{BraTS2021} & \multirow{3}{*}{\makecell{ \\ T1WI, T1CE\\T2WI, T2FLAIR}} & \cmark & \cmark & \cmark$^{*}$ & 644 & 92 & 184 \\
  \cmidrule(lr){2-2}   \cmidrule(lr){4-9}
& \makecell{BraTS-MEN} & & \cmark & \cmark & \cmark$^{*}$ & 644 & 92 & 184 \\
  \cmidrule(lr){2-2}   \cmidrule(lr){4-9}
& \makecell{BraTS-MET} & & \cmark & \cmark & \cmark$^{*}$ & 660 & 92 & 196 \\
  \cmidrule(lr){2-9}
& \makecell{ISLES2022} & \makecell{DWI, ADC} & \cmark & \cmark & \cmark$^{*}$ & 350 & 50 & 100 \\
  \cmidrule(lr){2-9}
& \makecell{WMH} & \makecell{T1WI \\ T2FLAIR}& \cmark & \cmark & \cmark$^{*}$ & 84 & 12 & 24 \\
\bottomrule
\end{tabular}
\label{tab:brainrg_summary}
\end{table}

\end{itemize}

\subsubsection*{Brain Structure Segmentation}

\begin{itemize}
    
\item[$\bullet$] \textbf{Hammers-n30r95~\cite{faillenot2017macroanatomy}}: This is a brain atlases dataset consisting of 30 adult T1-weighted MRIs, including annotations for 95 distinct brain anatomical regions\footnote{\url{http://brain-development.org/brain-atlases/}}. The dataset is utilized to quantify the brain structure segmentation performance of \OURMODEL. We randomly divided the dataset into training, validation, and testing sets with a ratio of 10:2:3. Additionally, we employed this dataset to generate brain structure segmentation labels for healthy Brain MRIs from our \OURDATASET~by performing registration.

\end{itemize}

\subsubsection*{Anomaly Segmentation}

\begin{itemize}
\setlength\itemsep{0.5em}
\item[$\bullet$] \textbf{BraTS~\cite{moawad2023brain,labella2023asnr,bakas2017segmentation,bakas2017advancing,menze2014multimodal,baid2021rsna,kazerooni2023brain,adewole2023brain}}: BraTS stands for Brain Tumor Segmentation (BraTS) Challenge. In this paper, we have used a series of datasets from BraTS, including BraTS2021, BraTS-PED, BraTS-SSA, BraTS-MET, and BraTS-MEN. Each patient data in the above datasets has 4 structural MRI volumes, including T1-weighted, post-contrast T1-weighted, T2-weighted, and T2 Fluid Attenuated Inversion Recovery, and associated manually generated ground truth labels for each tumor sub-region, including the necrotic (NCR) sections, the peritumoral edematous/invaded tissue (ED), and the enhancing tumor (ET). The details for each dataset are as follows: (i) BraTS2021 is a brain glioblastoma segmentation dataset consisting of data from 1,251 patients, which was all used to test our zero-shot anomaly segmentation performance after the first training stage of the segmentation module. Additionally, we randomly select 100 images comprising T1-weighted, T2-weighted, and T2-FLAIR modalities for assessing zero-shot brain structure segmentation performance. It is further used to train, validate, and test our report generation module as part of our \RELEASERG~dataset; (ii) BraTS-PED is used to train our segmentation module at the second stage with 79 cases of 237 images as training set and 20 cases of 60 images as a validation set; (iii) BraTS-SSA is used to train our segmentation module at the second stage with 48 cases of 144 images as training set and 12 cases of 36 images as a validation set; (iv) BraTS-MET is used to train our segmentation module at the second stage with 190 cases of 570 images as training set and 48 cases of 144 images as a validation set, and train, validate, and test our report generation module as part of our \RELEASERG~dataset; (v) BraTS-MEN is used to train our segmentation module at the second stage with 800 cases of 2,400 images as training set and 200 cases of 600 images as validation set, and train, validate, and test our report generation module as part of our \RELEASERG~dataset.

\item[$\bullet$] \textbf{ISLES2022~\cite{hernandez2022isles}}: 
This is an ischemic stroke lesion segmentation dataset consisting of 400 multi-vendor MRI cases with high variability in stroke lesion size, quantity, and location. In the official split, the dataset was split into a training dataset~(n=250) and a test dataset~(n=150). 
Since the test set is not released to the public, we use the training set of DWI modality to test our zero-shot anomaly segmentation performance after the first training stage of the segmentation module. Additionally, we randomly select 100 images of DWI to assess the zero-shot brain structure segmentation performance. It is further used to train and validate our segmentation module in the second stage and train, validate, and test our report generation module as part of our \RELEASERG~dataset.

\item[$\bullet$] \textbf{ATLAS~\cite{liew2017anatomical}}: ATLAS stands for anatomical tracings of lesions after stroke dataset. It has 655 patient data of T1WI modality with manually-segmented lesion masks. We use 524 cases of 524 images as training set and 131 cases of 131 images as validation set to train our segmentation module at the second stage.

\item[$\bullet$] \textbf{WMH~\cite{kuijf2019standardized}}: This is a challenge dataset for white matter hyperintensity (WMH) segmentation. It has 60 patient data of T1WI and T2FLAIR modalities with manually-segmented lesion masks from three institutions: UMC Utrecht, NUHS Singapore, and VU Amsterdam. The dataset is used to train and validate our segmentation module in the second stage with 48 cases of 96 images as train set and 12 cases of 24 images as validation set. Additionally, it is used to train, validate, and test our report generation module as part of our \RELEASERG~dataset.

\end{itemize}

\subsubsection*{Other Dataset}

\begin{itemize}
\item[$\bullet$] \textbf{RP3D-Brain}~\cite{wu2023towards}: 
RP3D-Brain is a subset of the RP3D dataset, 
with a focus on Brain MRI. The dataset is sourced from Radiopaedia\footnote{\url{https://radiopaedia.org/}}, an expanding peer-reviewed educational radiology resource website that enables clinicians to upload 3D volumes, providing a more realistic representation of clinical scenarios. Privacy concerns have been addressed by the clinicians during the uploading process. RP3D-Brain comprises 2,286 cases, encompassing 6,746 images from T1WI, T2WI, T2FLAIR, and DWI imaging modalities. Here, we use a subset of 100 cases from RP3D-Brain data to test the brain structure segmentation ability on abnormal brain scans of our segmentation module. 
\end{itemize}

\subsection{Evaluation}
Our proposed system, \OURMODELWOS, consists of two modules, serving three key functions: anomaly segmentation, brain anatomical structure segmentation, and report generation. Tab.~\ref{tab:baseline_summary} shows the organization of the evaluation, including the purpose, baseline methods, the test benchmark and their corresponding size.

\vspace{2pt} \noindent \textbf{Evaluation On Segmentation.}
We conduct a series of experiments to demonstrate the efficacy of our segmentation module on both internal and external data. The evaluation of the segmentation module is from \textbf{three aspects}: 
(i) anomaly segmentation on public datasets with quantitative results; (ii) brain structure segmentation on the public dataset with quantitative results; (iii) brain structure segmentation on abnormal brain MRIs with human ranking.
 
\vspace{2pt} \noindent \textbf{Evaluation On Report Generation.}
We evaluate the report generation on two distinct datasets: 
\OURDATASETWOS, an in-house dataset collected from the hospital, and \RELEASERGWOS, our collected grounded report generation dataset based on the public brain MRI segmentation samples.
To assess the quality of the generated reports, 
we compare them to the ground truth reports written and verified by senior doctors with 6-8 years of experience. 
For quantitative evaluation, we adopt popular metrics from the NLP community, for example, BLEU~\cite{papineni2002bleu}, ROUGE~\cite{lin2004rouge}, BERT-Score~\cite{zhang2019bertscore} and those specifically designed for measuring the radiology reports such as RadGraph~\cite{yu2023evaluating}, RadCliQ~\cite{yu2023evaluating}, and RateScore~\cite{zhao2024ratescore}. 
To demonstrate the system's practical usefulness in clinical scenarios, we conduct a human evaluation for generated reports on the \OURDATASET~dataset, which provided valuable insights into the generated reports' usability and effectiveness from the radiologists' perspective.

\vspace{2pt} \noindent \textbf{Integration in the Real Clinical Scenario.}
We integrate \OURMODEL~into the clinical workflow of one group of radiologists and compare it with the other group without AI assistance. Both groups of radiologists have 2-5 years of experience, 
and the ground truth reports are generated by senior radiologists with 6-8 years experience. Detailed analysis is presented in Sec.~\ref{sec:discussion} with specific cases listed in the Appendix Sec.\ref{sec:case_ai_assisted}.

\begin{table}[!t]
\centering
\caption{\textbf{Evaluation Overview}. 
The \OURDATASET~dataset is an in-house pre-training dataset collected from Shanghai Sixth People's Hospital, 
\RELEASERG~is our curated grounded report generation dataset, 
that will be released to the community.}
\vspace{3pt}
\renewcommand{\arraystretch}{1.1}
\footnotesize
\resizebox{1.0\linewidth}{!}{
\begin{tabular}{cc|c|c|c|c}
\toprule
\multicolumn{2}{c|}{Task} & Method & \makecell{Experiment \\ Setting}  & Test Benchmark & \makecell{\# Images}  \\
\midrule
\multicolumn{1}{c|}{\multirow{6}{*}{\makecell{\\ \\ \\ \\ \\ \\ \\ Module\\ Evaluation}}} & \multirow{1}{*}{\makecell{ Anomaly \\ Segmentation}} & \makecell{AE~\cite{ballard1987modular} VAE~\cite{kingma2013auto} VQ-VAE~\cite{van2017neural}\\ f-AnoGANE~\cite{schlegl2019f} BTAE~\cite{ghorbel2022transformer}  \\DCTAE~\cite{ghorbel2022transformer} HTAES~\cite{ghorbel2022transformer} CUAS~\cite{silva2022constrained}\\3D IF~\cite{naval2021implicit} Sim2Real~\cite{zhang2023self} \OURMODEL~(Ours)} & \makecell{Quantitative\\ Metrics} & \makecell{BraTS2021~\cite{baid2021rsna} \\ ISLES2022~\cite{hernandez2022isles}} & \makecell{2,051 \\ 247} \\
\cmidrule(lr){2-6}
\multicolumn{1}{c|}{} & \multirow{2}{*}{\makecell{\\ Brain \\ Structure\\ Segmentation}} & \multirow{2}{*}{\makecell{EasyReg~\cite{iglesias2023ready}\\ nnU-Net~\cite{isensee2021nnu}\\VoxHRNet~\cite{li2021whole}  \\ \OURMODEL~(Ours) \\Swin-UNETR~\cite{hatamizadeh2021swin}}}& \makecell{Quantitative\\ Metrics} & Hammers-n30r95~\cite{faillenot2017macroanatomy} & 6 \\
\cmidrule(lr){4-6}
\multicolumn{1}{c|}{} & &  & \makecell{Human \\ Evaluation }& \makecell{\OURDATASET~(Ours) \\ BraTS2021~\cite{baid2021rsna} \\ ISLES2022~\cite{hernandez2022isles} \\RP3D-Brain~\cite{wu2023towards}\\} & \makecell{100 \\ 100\\100\\100} \\
\cmidrule(lr){2-6}
\multicolumn{1}{c|}{} & \multirow{2}{*}{\makecell{ \\ Report \\ Generation}} & \makecell{Med-Flamingo~\cite{moor2023med} MedVInT~\cite{zhang2023pmc}\\ OpenFlamingo~\cite{awadalla2023openflamingo} \OURMODEL~(Ours)} & \makecell{Quantitative\\ Metrics} & \makecell{\OURDATASET~(Ours) \\ \RELEASERG~(Ours)} & \makecell{4,025 \\ 688} \\
\cmidrule(lr){3-6}
\multicolumn{1}{c|}{} & & \OURMODEL~(Ours) & \makecell{Human\\ Evaluation} &  \OURDATASET~(Ours) & 800 \\
\midrule
\multicolumn{2}{c|}{\makecell{Overall System \\ Evaluation}} & \OURMODEL~(Ours) & \makecell{AI-assisted \\ Report\\ Writing} &  \OURDATASET~(Ours) & 800  \\
\bottomrule
\end{tabular}}
\label{tab:baseline_summary}
\end{table}

\noindent \textbf{Baselines.} Tab.~\ref{tab:baseline_summary} lists the compared baselines. As \OURMODEL~is the first system capable of performing multiple tasks, such as anomaly segmentation, anatomical segmentation, and grounded report generation on brain MRI, here, we conduct comparisons with multiple existing approaches focusing on per task. The detailed introduction for each method is as follows: 

\subsubsection*{Anomaly Segmentation}
\begin{itemize}
\setlength\itemsep{0.5em}
\item \textbf{AE~\cite{ballard1987modular}}: An auto-encoder with a dense bottleneck is trained to reconstruct normal data, under the assumption that anomalies cannot be reconstructed as effectively.
\item \textbf{VAE~\cite{kingma2013auto}}: This method introduces a simple prior distribution for latent variables and uses KL-Divergence to constrain the latent space, differing from AE in its approach to handling the latent space.
\item \textbf{VQ-VAE~\cite{van2017neural}}: This method is a type of variational autoencoder that uses vector quantization to obtain a discrete latent representation. It differs from VAEs in two key ways: the encoder network outputs discrete, rather than continuous, codes; and the prior is learned rather than static. 
\item \textbf{f-AnoGANE~\cite{schlegl2019f}}: This model introduces a Generative Adversarial Network (GAN)-based unsupervised learning approach capable of identifying anomalous images and image segments, that can serve as an imaging biomarker candidates. 
\item \textbf{BTAE~\cite{ghorbel2022transformer}}: Basic transformer autoencoder is a simple autoencoder with transformer as the backbone.
\item \textbf{DCTAE~\cite{ghorbel2022transformer}}: Dense convolutional transformer autoencoder is a transformer-based AE model with a dense bottleneck.
\item \textbf{HTAES~\cite{ghorbel2022transformer}}: Hierarchical transformer autoencoder is a hierarchical transformer with
and without skip connections respectively, for gradually learning hierarchical features while reducing the computation complexity of standard Transformers for the medical image segmentation task.
\item \textbf{CUAS~\cite{silva2022constrained}}  Constrained unsupervised anomaly segmentation proposes a novel constrained formulation for unsupervised lesion segmentation, which integrates an auxiliary constrained loss to force the network to generate attention maps that cover the whole context in normal
images.
\item \textbf{3D IF~\cite{naval2021implicit}}: This method utilizes an auto-decoder to learn the distribution of healthy images, using implicit representation for mapping spatial coordinates to voxel intensities, with anomalies identified through restoration discrepancies.
\item  \textbf{Sim2Real~\cite{zhang2023self}}: This method is prior state-of-the-art self-supervised representation learning method for zero-shot tumor segmentation on 3D medical data, which was first pre-train a model with simulated tumors, and then adopt a self-training strategy for downstream data adaptation. We utilized the released checkpoint for zero-shot brain anomaly segmentation evaluation in our study. 
\end{itemize}

\subsubsection*{Brain Strucutre Segmentation}
\begin{itemize}
\setlength\itemsep{0.15cm}
    \item \textbf{EasyReg~\cite{iglesias2023ready}}: The EasyReg baseline refers to the SOTA registration method integrated in FreeSurfer 7.4. The segmentation results of EasyReg on the test set are obtained through image registration from the training set.
    \item \textbf{Swin-UNETR~\cite{hatamizadeh2021swin}}: Swin-UNETR is an architecture for semantic segmentation of brain tumors using multi-modal MRI images, which has a U-shaped network design and uses a Swin transformer as the encoder and CNN-based decoder that is connected to the encoder via skip connections at different resolutions. 
    \item \textbf{VoxHRNet~\cite{li2021whole}}: VoxHRNet is a full-volume framework, which feeds the full-volume brain image into the segmentation network and directly outputs the segmentation result for the whole brain volume.
    \item \textbf{nnU-Net~\cite{isensee2021nnu}}: nnU-Net is a SOTA deep learning-based segmentation method that automatically configures itself, including preprocessing, network srchitecture, training, and post-processing for any new task. Note the baseline nnU-Net is not the same structure as our segmentation module, since we modify the original nnU-Net to fit our problem scenario.
\end{itemize}

\subsubsection*{Report Generation}
The textual prompts for baselines under two types of evaluation are shown in Appendix Sec.\ref{sec:prompt_for_baselines}.
\begin{itemize}
\setlength\itemsep{0.15cm}
    \item  \textbf{OpenFlamingo~\cite{awadalla2023openflamingo}}: This is an open-source implementation of the prior state-of-the-art generalist visual language model Flamingo~\cite{alayrac2022flamingo}, which was trained on large-scale data from a general visual-language domain. We utilized the released checkpoint for \textbf{zero-shot} and \textbf{few-shot} evaluation in our study. 
    
    \item  \textbf{MedVInT~\cite{zhang2023pmc}}: This is a visual instruction-tuned visual language model based on LLaMA~\cite{touvron2023llama}, which was trained on PMC-VQA~\cite{zhang2023pmc}. Considering that the PMC-VQA data does not contain any few-shot cases, mainly targeting at zero-shot prompting cases, we directly use the released checkpoint of the MedVInT-TD model with PMC-LLaMA and PMC-CLIP backbone for \textbf{zero-shot} evaluation.

    \item  \textbf{Med-Flamingo~\cite{moor2023med}}: This is a multimodal model developed based on OpenFlamingo-9B~\cite{awadalla2023openflamingo}, that can handles multi-image input interleaving with texts. We use the released checkpoint for \textbf{zero-shot} and \textbf{few-shot} evaluation. 
\end{itemize}

\noindent \textbf{Metrics.} We adopt various metrics to assess our system under different evaluation tasks. Specifically, For the segmentation tasks, we quantitatively evaluate the performance from the perspective of Dice Similarity Coefficient~(DSC), Precision (PRE), and Sensitivity (SE) respectively.
For the report generation tasks, we evaluate the generative models with BLEU-4, ROUGE-1, BERT-Score, RadGraph, RadCliQ, and RaTEScore scores.
The detailed introduction of each metric is as follows:

\begin{itemize}
\setlength\itemsep{0.15cm}

\item \textbf{DSC.} Dice Similarity Coefficient~(DSC) is a standard \textbf{segmentation} metric. It measures the overlap between the model's prediction $P$ and ground truth $G$, formally defined as:
\begin{equation}
    DSC(P,G) = \frac{2|P \bigcap G|}{|P|+|G|}.
    \nonumber
\end{equation}

\item \textbf{PRE.} Precision~(PRE)

is a \textbf{segmentation} metric that measures the probability of correct prediction, which is defined as:
\begin{equation}
    PRE(P,G) = \frac{|P \bigcap G|}{|P \bigcap G|+|P - G|}.
    \nonumber
\end{equation}

\item \textbf{SE.} Sensitivity~(SE)

is a \textbf{segmentation} metric that measures the model's ability to correctly detect patients who have the disease, which is defined as:
\begin{equation}
    SE(P,G) = \frac{|P \bigcap G|}{|P \bigcap G|+|G - P|}.
    \nonumber
\end{equation}

\item \textbf{BLEU~\cite{papineni2002bleu}.} BLEU stands for ``BiLingual Evaluation Understudy''

It measures the accuracy of the \textbf{report generation} result. We use \texttt{nltk.translate.bleu\_score.sentence\_bleu} with weights of $(0,0,0,1)$ to calculate our 4-gram BLEU-4 score.

\item \textbf{ROUGE~\cite{lin2004rouge}.} ROUGE stands for ``Recall-Oriented Understudy for Gisting Evaluation''. It refers to a set of \textbf{report generation} evaluation measures that assess the overlap between a generated summary and a set of reference summaries. Assume the reference report as $G$, the $\text{ROUGE}_{n}$ metric ($n=1$ by default) is calculated as follows:
\begin{equation}
    ROUGE_{n} = \frac{\sum_{s\in G}\sum_{gram_{n}\in s}Count_{match}(gram_{n})}{\sum_{c\in G}\sum_{gram_{n}\in s}Count(gram_{n})},
    \nonumber
\end{equation}
here we use the $\text{ROUGE}_{1}$ as our ROUGE-1 metric.

\item \textbf{BERT-Score~\cite{zhang2019bertscore}.} BERTScore is a \textbf{report generation} metric that measures the \textbf{report generation} similarity between the generated report and the reference report using contextualized word embeddings provided by the BERT (Bidirectional Encoder Representations from Transformers) model. In this paper, we use \texttt{bert\_score.BERT-Scorer} with model type as \texttt{bert-base-uncased} to calculate the BERT-Score.

\item \textbf{RadGraph~\cite{yu2023evaluating}.} An automatic \textbf{report generation} metric that computes the overlap in clinical entities and relations between a model-generated report and a radiologist-generated report. For simplicity, we use RadGraph here to represent the `RadGaph F1' in the original paper.

\item \textbf{RadCliQ~\cite{yu2023evaluating}.} This \textbf{report generation} metric predicts a radiologist-determined error score from a combination of automated metrics, including BLEU, BERTScore, CheXbert vector similarity, and RadGraph.

\item \textbf{RaTEScore~\cite{zhao2024ratescore}.}  RaTEScore is a \textbf{report generation} metric that extracts key medical terms and computes the similarity based on the entity embeddings computed from the language model and their corresponding types. It is robust against complex medical synonyms and sensitive to negation expressions.
\end{itemize}

\section{Conclusion}
In this paper, we propose the first brain MRI interpretation system, termed \OURMODELWOS, capable of analyzing brain MRIs through brain anomaly segmentation, brain structure segmentation, and report generation, closely following radiologists' workflow. We also build \RELEASERGWOS, a new benchmark dataset for grounded report generation on brain MRI with anomaly segmentation and report annotations, encompassing 3,408 images across 6 modalities and 5 disease types. 
Experimentally, we demonstrate the superior performance of each system sub-module and the effectiveness of \RELEASERG~as a pre-training dataset.  
Lastly, we show that integrating \OURMODEL~into clinician workflows can provide educational benefits for junior radiologists and enhance overall work efficiency. 
We open-source \OURMODELWOS, \RELEASERGWOS, and our human evaluation platform to boost future research.

\bibliographystyle{sn-mathphys} 
\bibliography{sn-bibliography} 

\begin{thebibliography}{10}\itemsep=-1pt

\bibitem{achiam2023gpt}
Josh Achiam, Steven Adler, Sandhini Agarwal, Lama Ahmad, Ilge Akkaya, Florencia~Leoni Aleman, Diogo Almeida, Janko Altenschmidt, Sam Altman, Shyamal Anadkat, et~al.
\newblock Gpt-4 technical report.
\newblock {\em arXiv preprint arXiv:2303.08774}, 2023.

\bibitem{adewole2023brain}
Maruf Adewole, Jeffrey~D Rudie, Anu Gbdamosi, Oluyemisi Toyobo, Confidence Raymond, Dong Zhang, Olubukola Omidiji, Rachel Akinola, Mohammad~Abba Suwaid, Adaobi Emegoakor, et~al.
\newblock The brain tumor segmentation (brats) challenge 2023: glioma segmentation in sub-saharan africa patient population (brats-africa).
\newblock {\em ArXiv}, 2023.

\bibitem{alayrac2022flamingo}
Jean-Baptiste Alayrac, Jeff Donahue, Pauline Luc, Antoine Miech, Iain Barr, Yana Hasson, Karel Lenc, Arthur Mensch, Katherine Millican, Malcolm Reynolds, et~al.
\newblock Flamingo: a visual language model for few-shot learning.
\newblock {\em Advances in neural information processing systems}, 35:23716--23736, 2022.

\bibitem{awadalla2023openflamingo}
Anas Awadalla, Irena Gao, Josh Gardner, Jack Hessel, Yusuf Hanafy, Wanrong Zhu, Kalyani Marathe, Yonatan Bitton, Samir Gadre, Shiori Sagawa, et~al.
\newblock Openflamingo: An open-source framework for training large autoregressive vision-language models.
\newblock {\em arXiv preprint arXiv:2308.01390}, 2023.

\bibitem{baid2021rsna}
Ujjwal Baid, Satyam Ghodasara, Suyash Mohan, Michel Bilello, Evan Calabrese, Errol Colak, Keyvan Farahani, Jayashree Kalpathy-Cramer, Felipe~C Kitamura, Sarthak Pati, et~al.
\newblock The rsna-asnr-miccai brats 2021 benchmark on brain tumor segmentation and radiogenomic classification.
\newblock {\em arXiv preprint arXiv:2107.02314}, 2021.

\bibitem{bakas2017segmentation}
Spyridon Bakas, Hamed Akbari, Aristeidis Sotiras, Michel Bilello, Martin Rozycki, Justin Kirby, John Freymann, Keyvan Farahani, and Christos Davatzikos.
\newblock Segmentation labels and radiomic features for the pre-operative scans of the tcga-lgg collection.
\newblock {\em The cancer imaging archive}, 286, 2017.

\bibitem{bakas2017advancing}
Spyridon Bakas, Hamed Akbari, Aristeidis Sotiras, Michel Bilello, Martin Rozycki, Justin~S Kirby, John~B Freymann, Keyvan Farahani, and Christos Davatzikos.
\newblock Advancing the cancer genome atlas glioma mri collections with expert segmentation labels and radiomic features.
\newblock {\em Scientific data}, 4(1):1--13, 2017.

\bibitem{ballard1987modular}
Dana~H Ballard.
\newblock Modular learning in neural networks.
\newblock In {\em Proceedings of the Sixth National Conference on Artificial Intelligence}, pages 279--284, 1987.

\bibitem{chen2020generating}
Zhihong Chen, Yan Song, Tsung-Hui Chang, and Xiang Wan.
\newblock Generating radiology reports via memory-driven transformer.
\newblock In {\em Proceedings of the 2020 Conference on Empirical Methods in Natural Language Processing (EMNLP)}, pages 1439--1449, 2020.

\bibitem{demner2016preparing}
Dina Demner-Fushman, Marc~D Kohli, Marc~B Rosenman, Sonya~E Shooshan, Laritza Rodriguez, Sameer Antani, George~R Thoma, and Clement~J McDonald.
\newblock Preparing a collection of radiology examinations for distribution and retrieval.
\newblock {\em Journal of the American Medical Informatics Association}, 23(2):304--310, 2016.

\bibitem{faillenot2017macroanatomy}
Isabelle Faillenot, Rolf~A Heckemann, Maud Frot, and Alexander Hammers.
\newblock Macroanatomy and 3d probabilistic atlas of the human insula.
\newblock {\em Neuroimage}, 150:88--98, 2017.

\bibitem{ghorbel2022transformer}
Ahmed Ghorbel, Ahmed Aldahdooh, Shadi Albarqouni, and Wassim Hamidouche.
\newblock Transformer based models for unsupervised anomaly segmentation in brain mr images.
\newblock In {\em International MICCAI Brainlesion Workshop}, pages 25--44. Springer, 2022.

\bibitem{hatamizadeh2021swin}
Ali Hatamizadeh, Vishwesh Nath, Yucheng Tang, Dong Yang, Holger~R Roth, and Daguang Xu.
\newblock Swin unetr: Swin transformers for semantic segmentation of brain tumors in mri images.
\newblock In {\em International MICCAI Brainlesion Workshop}, pages 272--284. Springer, 2021.

\bibitem{hernandez2022isles}
Moritz~R Hernandez~Petzsche, Ezequiel de~la Rosa, Uta Hanning, Roland Wiest, Waldo Valenzuela, Mauricio Reyes, Maria Meyer, Sook-Lei Liew, Florian Kofler, Ivan Ezhov, et~al.
\newblock Isles 2022: A multi-center magnetic resonance imaging stroke lesion segmentation dataset.
\newblock {\em Scientific data}, 9(1):762, 2022.

\bibitem{hu2023label}
Qixin Hu, Yixiong Chen, Junfei Xiao, Shuwen Sun, Jieneng Chen, Alan~L Yuille, and Zongwei Zhou.
\newblock Label-free liver tumor segmentation.
\newblock In {\em Proceedings of the IEEE/CVF Conference on Computer Vision and Pattern Recognition}, pages 7422--7432, 2023.

\bibitem{iglesias2023ready}
Juan~Eugenio Iglesias.
\newblock A ready-to-use machine learning tool for symmetric multi-modality registration of brain mri.
\newblock {\em Scientific Reports}, 13(1):6657, 2023.

\bibitem{isensee2021nnu}
Fabian Isensee, Paul~F Jaeger, Simon~AA Kohl, Jens Petersen, and Klaus~H Maier-Hein.
\newblock nnu-net: a self-configuring method for deep learning-based biomedical image segmentation.
\newblock {\em Nature methods}, 18(2):203--211, 2021.

\bibitem{johnson2019mimic}
Alistair~EW Johnson, Tom~J Pollard, Seth~J Berkowitz, Nathaniel~R Greenbaum, Matthew~P Lungren, Chih-ying Deng, Roger~G Mark, and Steven Horng.
\newblock Mimic-cxr, a de-identified publicly available database of chest radiographs with free-text reports.
\newblock {\em Scientific data}, 6(1):317, 2019.

\bibitem{kazerooni2023brain}
Anahita~Fathi Kazerooni, Nastaran Khalili, Xinyang Liu, Debanjan Haldar, Zhifan Jiang, Syed~Muhammed Anwar, Jake Albrecht, Maruf Adewole, Udunna Anazodo, Hannah Anderson, et~al.
\newblock The brain tumor segmentation (brats) challenge 2023: Focus on pediatrics (cbtn-connect-dipgr-asnr-miccai brats-peds).
\newblock {\em arXiv preprint arXiv:2305.17033}, 2023.

\bibitem{kingma2013auto}
Diederik~P Kingma and Max Welling.
\newblock Auto-encoding variational bayes.
\newblock {\em arXiv preprint arXiv:1312.6114}, 2013.

\bibitem{kuijf2019standardized}
Hugo~J Kuijf, J~Matthijs Biesbroek, Jeroen De~Bresser, Rutger Heinen, Simon Andermatt, Mariana Bento, Matt Berseth, Mikhail Belyaev, M~Jorge Cardoso, Adria Casamitjana, et~al.
\newblock Standardized assessment of automatic segmentation of white matter hyperintensities and results of the wmh segmentation challenge.
\newblock {\em IEEE transactions on medical imaging}, 38(11):2556--2568, 2019.

\bibitem{labella2023asnr}
Dominic LaBella, Maruf Adewole, Michelle Alonso-Basanta, Talissa Altes, Syed~Muhammad Anwar, Ujjwal Baid, Timothy Bergquist, Radhika Bhalerao, Sully Chen, Verena Chung, et~al.
\newblock The asnr-miccai brain tumor segmentation (brats) challenge 2023: Intracranial meningioma.
\newblock {\em arXiv preprint arXiv:2305.07642}, 2023.

\bibitem{li2023dynamic}
Mingjie Li, Bingqian Lin, Zicong Chen, Haokun Lin, Xiaodan Liang, and Xiaojun Chang.
\newblock Dynamic graph enhanced contrastive learning for chest x-ray report generation.
\newblock In {\em Proceedings of the IEEE/CVF Conference on Computer Vision and Pattern Recognition}, pages 3334--3343, 2023.

\bibitem{li2021whole}
Yeshu Li, Jonathan Cui, Yilun Sheng, Xiao Liang, Jingdong Wang, I Eric, Chao Chang, and Yan Xu.
\newblock Whole brain segmentation with full volume neural network.
\newblock {\em Computerized Medical Imaging and Graphics}, 93:101991, 2021.

\bibitem{liew2017anatomical}
Sook-Lei Liew, Julia~M Anglin, Nick~W Banks, Matt Sondag, Kaori~L Ito, Hosung Kim, Jennifer Chan, Joyce Ito, Connie Jung, Stephanie Lefebvre, et~al.
\newblock The anatomical tracings of lesions after stroke (atlas) dataset-release 1.1.
\newblock {\em bioRxiv}, page 179614, 2017.

\bibitem{lin2004rouge}
Chin-Yew Lin.
\newblock Rouge: A package for automatic evaluation of summaries.
\newblock In {\em Text Summarization Branches Out}, pages 74--81, 2004.

\bibitem{liu2019clinically}
Guanxiong Liu, Tzu-Ming~Harry Hsu, Matthew McDermott, Willie Boag, Wei-Hung Weng, Peter Szolovits, and Marzyeh Ghassemi.
\newblock Clinically accurate chest x-ray report generation.
\newblock In {\em Machine Learning for Healthcare Conference}, pages 249--269. PMLR, 2019.

\bibitem{giles2019error}
Giles Maskell and FRCR FRCP.
\newblock Error in radiology—where are we now?
\newblock {\em Br J Radiol}, 92:1095, 2019.

\bibitem{mcdonald2015effects}
Robert~J McDonald, Kara~M Schwartz, Laurence~J Eckel, Felix~E Diehn, Christopher~H Hunt, Brian~J Bartholmai, Bradley~J Erickson, and David~F Kallmes.
\newblock The effects of changes in utilization and technological advancements of cross-sectional imaging on radiologist workload.
\newblock {\em Academic radiology}, 22(9):1191--1198, 2015.

\bibitem{menze2014multimodal}
Bjoern~H Menze, Andras Jakab, Stefan Bauer, Jayashree Kalpathy-Cramer, Keyvan Farahani, Justin Kirby, Yuliya Burren, Nicole Porz, Johannes Slotboom, Roland Wiest, et~al.
\newblock The multimodal brain tumor image segmentation benchmark (brats).
\newblock {\em IEEE transactions on medical imaging}, 34(10):1993--2024, 2014.

\bibitem{moawad2023brain}
Ahmed~W Moawad, Anastasia Janas, Ujjwal Baid, Divya Ramakrishnan, Leon Jekel, Kiril Krantchev, Harrison Moy, Rachit Saluja, Klara Osenberg, Klara Wilms, et~al.
\newblock The brain tumor segmentation (brats-mets) challenge 2023: Brain metastasis segmentation on pre-treatment mri.
\newblock {\em ArXiv}, 2023.

\bibitem{moor2023foundation}
Michael Moor, Oishi Banerjee, Zahra Shakeri~Hossein Abad, Harlan~M Krumholz, Jure Leskovec, Eric~J Topol, and Pranav Rajpurkar.
\newblock Foundation models for generalist medical artificial intelligence.
\newblock {\em Nature}, 616(7956):259--265, 2023.

\bibitem{moor2023med}
Michael Moor, Qian Huang, Shirley Wu, Michihiro Yasunaga, Yash Dalmia, Jure Leskovec, Cyril Zakka, Eduardo~Pontes Reis, and Pranav Rajpurkar.
\newblock Med-flamingo: a multimodal medical few-shot learner.
\newblock In {\em Machine Learning for Health (ML4H)}, pages 353--367. PMLR, 2023.

\bibitem{naval2021implicit}
Sergio Naval~Marimont and Giacomo Tarroni.
\newblock Implicit field learning for unsupervised anomaly detection in medical images.
\newblock In {\em Medical Image Computing and Computer Assisted Intervention--MICCAI 2021: 24th International Conference, Strasbourg, France, September 27--October 1, 2021, Proceedings, Part II 24}, pages 189--198. Springer, 2021.

\bibitem{nooralahzadeh2021progressive}
Farhad Nooralahzadeh, Nicolas~Perez Gonzalez, Thomas Frauenfelder, Koji Fujimoto, and Michael Krauthammer.
\newblock Progressive transformer-based generation of radiology reports.
\newblock {\em arXiv preprint arXiv:2102.09777}, 2021.

\bibitem{papanikolaou2020dare}
Yannis Papanikolaou and Andrea Pierleoni.
\newblock Dare: Data augmented relation extraction with gpt-2.
\newblock {\em arXiv preprint arXiv:2004.13845}, 2020.

\bibitem{papineni2002bleu}
Kishore Papineni, Salim Roukos, Todd Ward, and Wei-Jing Zhu.
\newblock Bleu: a method for automatic evaluation of machine translation.
\newblock In {\em Proceedings of the 40th annual meeting of the Association for Computational Linguistics}, pages 311--318, 2002.

\bibitem{Poorter1999GrowthRO}
Lourens Poorter.
\newblock Growth responses of 15 rain‐forest tree species to a light gradient: the relative importance of morphological and physiological traits.
\newblock {\em Functional Ecology}, 13:396--410, 1999.

\bibitem{radford2019language}
Alec Radford, Jeffrey Wu, Rewon Child, David Luan, Dario Amodei, Ilya Sutskever, et~al.
\newblock Language models are unsupervised multitask learners.
\newblock {\em OpenAI blog}, 1(8):9, 2019.

\bibitem{schlegl2019f}
Thomas Schlegl, Philipp Seeb{\"o}ck, Sebastian~M Waldstein, Georg Langs, and Ursula Schmidt-Erfurth.
\newblock f-anogan: Fast unsupervised anomaly detection with generative adversarial networks.
\newblock {\em Medical image analysis}, 54:30--44, 2019.

\bibitem{schluter2022natural}
Hannah~M Schl{\"u}ter, Jeremy Tan, Benjamin Hou, and Bernhard Kainz.
\newblock Natural synthetic anomalies for self-supervised anomaly detection and localization.
\newblock In {\em European Conference on Computer Vision}, pages 474--489. Springer, 2022.

\bibitem{silva2022constrained}
Julio Silva-Rodr{\'\i}guez, Valery Naranjo, and Jose Dolz.
\newblock Constrained unsupervised anomaly segmentation.
\newblock {\em Medical Image Analysis}, 80:102526, 2022.

\bibitem{tan2021detecting}
Jeremy Tan, Benjamin Hou, Thomas Day, John Simpson, Daniel Rueckert, and Bernhard Kainz.
\newblock Detecting outliers with poisson image interpolation.
\newblock In {\em Medical Image Computing and Computer Assisted Intervention--MICCAI 2021: 24th International Conference, Strasbourg, France, September 27--October 1, 2021, Proceedings, Part V 24}, pages 581--591. Springer, 2021.

\bibitem{touvron2023llama}
Hugo Touvron, Thibaut Lavril, Gautier Izacard, Xavier Martinet, Marie-Anne Lachaux, Timoth{\'e}e Lacroix, Baptiste Rozi{\`e}re, Naman Goyal, Eric Hambro, Faisal Azhar, et~al.
\newblock Llama: Open and efficient foundation language models.
\newblock {\em arXiv preprint arXiv:2302.13971}, 2023.

\bibitem{tu2024towards}
Tao Tu, Shekoofeh Azizi, Danny Driess, Mike Schaekermann, Mohamed Amin, Pi-Chuan Chang, Andrew Carroll, Charles Lau, Ryutaro Tanno, Ira Ktena, et~al.
\newblock Towards generalist biomedical ai.
\newblock {\em NEJM AI}, 1(3):AIoa2300138, 2024.

\bibitem{van2017neural}
Aaron Van Den~Oord, Oriol Vinyals, et~al.
\newblock Neural discrete representation learning.
\newblock {\em Advances in neural information processing systems}, 30, 2017.

\bibitem{wu2023towards}
Chaoyi Wu, Xiaoman Zhang, Ya Zhang, Yanfeng Wang, and Weidi Xie.
\newblock Towards generalist foundation model for radiology.
\newblock {\em arXiv preprint arXiv:2308.02463}, 2023.

\bibitem{yu2023evaluating}
Feiyang Yu, Mark Endo, Rayan Krishnan, Ian Pan, Andy Tsai, Eduardo~Pontes Reis, Eduardo Kaiser Ururahy~Nunes Fonseca, Henrique Min~Ho Lee, Zahra Shakeri~Hossein Abad, Andrew~Y Ng, et~al.
\newblock Evaluating progress in automatic chest x-ray radiology report generation.
\newblock {\em Patterns}, 4(9), 2023.

\bibitem{zhang2019bertscore}
Tianyi Zhang, Varsha Kishore, Felix Wu, Kilian~Q Weinberger, and Yoav Artzi.
\newblock Bertscore: Evaluating text generation with bert.
\newblock {\em arXiv preprint arXiv:1904.09675}, 2019.

\bibitem{zhang2023unsupervised}
Xinru Zhang, Ni Ou, Chenghao Liu, Zhizheng Zhuo, Yaou Liu, and Chuyang Ye.
\newblock Unsupervised brain tumor segmentation with image-based prompts.
\newblock {\em arXiv preprint arXiv:2304.01472}, 2023.

\bibitem{zhang2024radgenome-chest-ct}
Xiaoman Zhang, Chaoyi Wu, Ziheng Zhao, Jiayu Lei, Ya Zhang, Yanfeng Wang, and Weidi Xie.
\newblock Radgenome-chest ct: A grounded vision-language dataset for chest ct analysis.
\newblock {\em arXiv preprint arXiv:2404.16754}, 2024.

\bibitem{zhang2023pmc}
Xiaoman Zhang, Chaoyi Wu, Ziheng Zhao, Weixiong Lin, Ya Zhang, Yanfeng Wang, and Weidi Xie.
\newblock Pmc-vqa: Visual instruction tuning for medical visual question answering.
\newblock {\em arXiv preprint arXiv:2305.10415}, 2023.

\bibitem{zhang2023self}
Xiaoman Zhang, Weidi Xie, Chaoqin Huang, Ya Zhang, Xin Chen, Qi Tian, and Yanfeng Wang.
\newblock Self-supervised tumor segmentation with sim2real adaptation.
\newblock {\em IEEE Journal of Biomedical and Health Informatics}, 2023.

\bibitem{zhao2024ratescore}
Weike Zhao, Chaoyi Wu, Xiaoman Zhang, Ya Zhang, Yanfeng Wang, and Weidi Xie.
\newblock Ratescore: A metric for radiology report generation.
\newblock {\em arXiv preprint arXiv:2406.16845}, 2024.

\bibitem{zhao2023model}
Ziheng Zhao, Yao Zhang, Chaoyi Wu, Xiaoman Zhang, Ya Zhang, Yanfeng Wang, and Weidi Xie.
\newblock One model to rule them all: Towards universal segmentation for medical images with text prompt, 2023.

\bibitem{zhou2021visual}
Yi Zhou, Lei Huang, Tao Zhou, Huazhu Fu, and Ling Shao.
\newblock Visual-textual attentive semantic consistency for medical report generation.
\newblock In {\em Proceedings of the IEEE/CVF International Conference on Computer Vision}, pages 3985--3994, 2021.

\end{thebibliography}

\clearpage
\appendix
\section{Segmentation Evaluation}

\captionsetup[figure]{name=Supplementary Fig.}
\captionsetup[table]{name=Supplementary Table }
\setcounter{figure}{0}
\setcounter{table}{0}

We adopt the Swin-UNETR, VoxHRNet, and nnU-Net as the most widely utilized segmentation backbones, alongside the state-of-the-art registration model EasyReg, serving as the models to verify our self-supervised regime. As we do not have the ground truth anatomical segmentation mask on multi-modal MRI scans of abnormal brains, we conduct the quantitative evaluation on synthetic data, by adding artificial `lesions'~\cite{zhang2023self,hu2023label,schluter2022natural,tan2021detecting}. 
We split the Hammer-b30r95 dataset into the same training, validation, and testing sets for all methods, except for EasyReg, a registration tool that does not require training. 
Results in Table~\ref{tab:brain_region_hammer} demonstrate that our self-supervised regime significantly improves brain region segmentation on lesion-induced brain across various network backbones. 

\begin{table}[!htb]
\centering
\small
\caption{ Effectiveness of our self-supervised training strategy on different widely-used network backbones. We evaluate these models (except registration model EasyReg) w./w.o our self-supervised strategy on two different datasets, the original Hammers-n30r95 dataset and the Hammers-n30r95 dataset with synthetic anomalies. We report the Dice Similarity Coefficient (DSC) (\%), Hausdorff Distance (HD), Precision (PRE) (\%) and Sensitivity (SE) (\%). The best result is bolded.}
\vspace{3pt}
\renewcommand{\arraystretch}{1.1}
\setlength{\tabcolsep}{{5pt}}
\begin{tabular}{lccccccccc}
\toprule
 \multirow{2}{*}{Network} & \multirow{2}{*}{\makecell{Signal-aware\\Synthetic Anomalies}} & \multicolumn{4}{c}{Hammers-n30r95} &\multicolumn{4}{c}{\makecell{Hammers-n30r95 \\
w. Synthetic Anomalies}} \\
\cmidrule(lr){3-6} \cmidrule(lr){7-10}
 & & DSC $\uparrow$ & HD $\downarrow$ & PRE $\uparrow$ & SE $\uparrow$ & DSC $\uparrow$ & HD $\downarrow$ & PRE $\uparrow$ & SE $\uparrow$ \\

\cmidrule(lr){1-10}
EasyReg~\cite{iglesias2023ready} & \usym{2718} & 69.03 & 10.69 & 69.56 & 70.78 & 67.64 & 10.79 & 68.6 & 69.02  \\
\cmidrule(lr){1-10}
Swin-UNETR~\cite{hatamizadeh2021swin} & \usym{2718} & 79.30 & 14.69 & 80.36 & 79.54 & 72.89 & 34.37 & 78.03 & 72.26 \\
Swin-UNETR~\cite{hatamizadeh2021swin} & \ding{52} & 77.65 & 25.88 & 78.69 & 78.17 & 74.55 & 29.04 & 76.47 & 74.71 \\
\cmidrule(lr){1-10}
VoxHRNet~\cite{li2021whole} & \usym{2718} & 80.32 & 8.68 & 79.96 & 81.74 & 74.53 & 14.97 & 76.41 & 74.94 \\
VoXHRNet~\cite{li2021whole} & \ding{52} & 80.34 & 8.65 & 79.95 & 81.84 & 77.15 & 11.18 & 80.13 & 79.49 \\
\cmidrule(lr){1-10}
nnU-Net~\cite{isensee2021nnu} & \usym{2718} & \textbf{81.85} & \textbf{8.86} & \textbf{81.96} & \textbf{82.85} & 75.19 & 13.74 & 77.24 & 75.62 \\
nnU-Net~\cite{isensee2021nnu} & \ding{52} & 81.52 & 9.16 & 81.68 & 82.63 & \textbf{79.57} & \textbf{9.28} & \textbf{81.43} & \textbf{81.57} \\
\bottomrule
\label{tab:brain_region_hammer}
\end{tabular}
\end{table}

\begin{figure}[!htb]
\centering
\includegraphics[width=\textwidth]{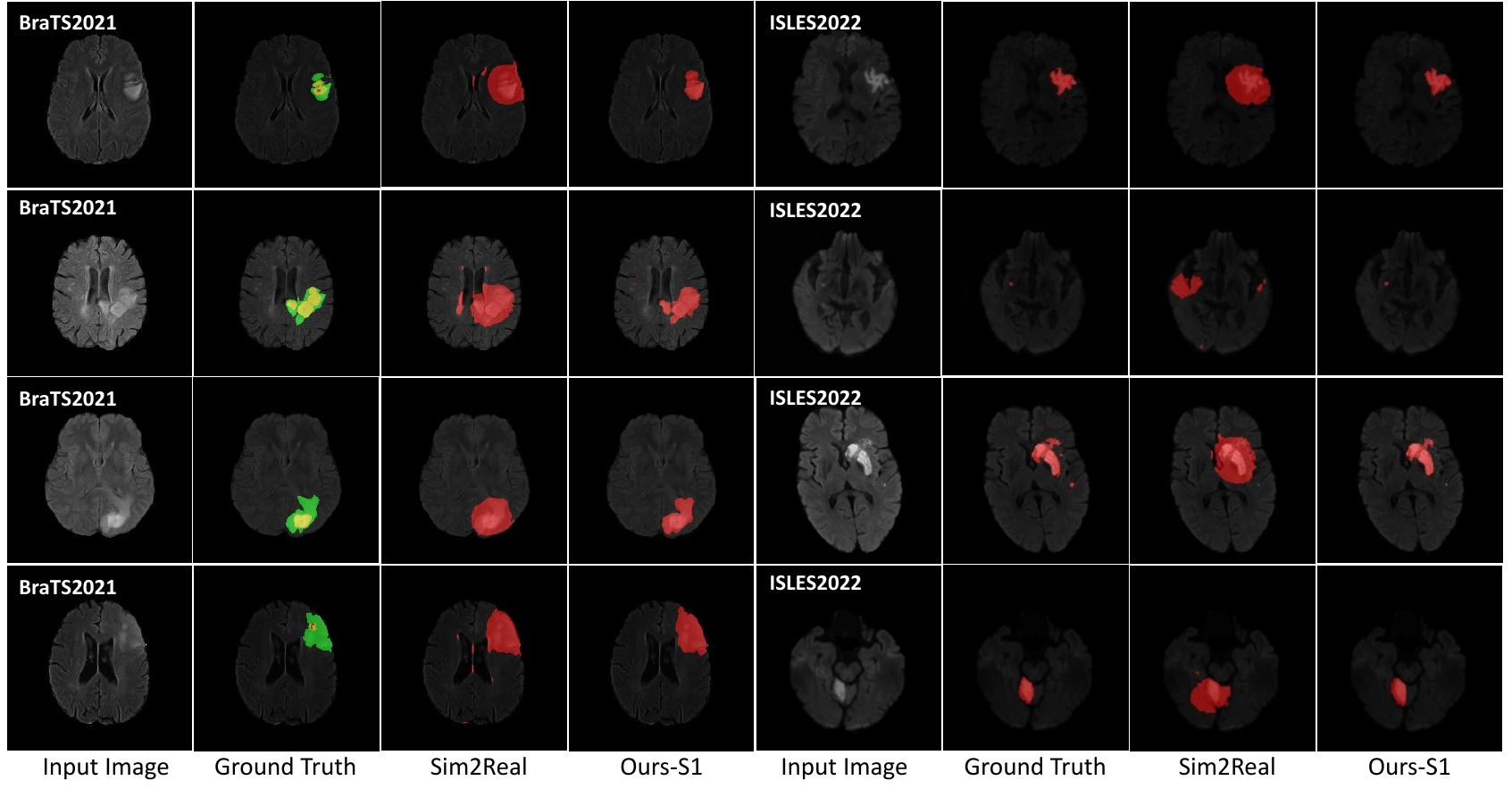}
\caption{Examples of abnormal segmentation of Sim2Real and Ours-S1, our segmentation module after the first self-supervised training stage, on the BraTS2021 and ISLES2022 datasets.}
\label{fig:anomaly_segmentation_example}
\end{figure}
\clearpage

\section{Report Generation Evaluation}

\subsection{Prompt for Report Generation Baselines.} \label{sec:prompt_for_baselines}
For OpenFlamingo and Med-Flamingo, we perform both zero-shot and few-shot evaluations in our study. Specifically, we follow the prompts derived from the official Med-Flamingo repository.
The zero-shot evaluation prompt for whole report generation is as follows: 
\begin{mdframed}[backgroundcolor=gray!10] 
`You are a helpful radiology assistant. You are being provided with an image and a question, answer the given question. <image> Question: Please write a radiology report that explains this brain MRI image. Answer:''.
\end{mdframed}

The zero-shot evaluation prompt for region-based report generation is as follows: 
\begin{mdframed}[backgroundcolor=gray!10] 
`You are a helpful radiology assistant. You are being provided with an image and a question, answer the given question. <image> Question: Please write a summarized radiology report that describe the findings/abnormality in the [region\_text] of this brain MRI image. Answer:''.
\end{mdframed}
where [region\_text] denotes the target region list.
In the few-shot setting, we expand upon this format by supplying the models with additional examples to guide their responses. 
This is structured as follows: 
\begin{mdframed}[backgroundcolor=gray!10] 
``You are a helpful medical assistant. You are being provided with images, a question about the image, and an answer. Follow the examples and answer the last question. $<$image$>$ Question: [the first question] Answer: [the first answer]. $<|$endofchunk$|>$ <image> Question: [the second question] Answer: [the second answer].<|endofchunk|> <image> Question: [the third question] Answer: [the third answer].<|endofchunk|> <image> Question: Please write a radiology report that explains this brain MRI image. Answer:''.
\end{mdframed}
\clearpage
\subsection{Prompt for Report Processing}\label{prompt_report_decomp}

\subsubsection{Prompt for \OURDATASET~Report Processing}

\begin{mdframed}[backgroundcolor=gray!10]

GPT-4, I need you to analyze the following MRI report and break it down into a structured JSON format. The JSON should organize the findings by anatomical regions and MRI sequences, with each related observation described in a complete sentence derieved from the original report. For each region and sequence, mention if the signal is isointense, hyperintense, etc. Here's the report: <report>. Please compile the findings in the following JSON format:\\
    \{\\
        'Region\_Name' \{\\
        'T1W' (if mentioned): 'Description of the lesion in the original reports on T1W sequence, including its boundary, edema, its impact to surrouding brain tissue (if any).',\\
        'T2W' (if mentioned): 'Description of the lesion in the original reports on T2W sequence, including its boundary, edema, its impact to surrouding brain tissue (if any).',\\
        'FLAIR' (if mentioned): 'Description of the lesion in the original reports on FLAIR sequence, including its boundary, edema, its impact to surrouding brain tissue (if any).',\\
        'DWI' (if mentioned): 'Description of the lesion in the original reports on DWI sequence, including its boundary, edema, its impact to surrouding brain tissue (if any).',\\
        'Other' (if not any sequence is mentioned): 'Description of the region in the original reports.'\\
        \},\\
    ...
    \}\\
    Begin when you're ready.
\end{mdframed}

\subsubsection{Prompt for BraTS2021/MET/MEN Report Processing}

\begin{mdframed}[backgroundcolor=gray!10]
GPT-4, I need you to analyze the following chinese MRI report fingdings and impression. For the finding:\\
1. translate the provided report findings into english with precise medical terminology. Ensure that the translation accurately conveys the original meaning and includes all the relevant medical terms.\\
2. Break the english fingdings down into a structured format. The JSON should organize the findings by anatomical regions and MRI sequences, with each related observation described in a complete sentence derieved from the original report. For each region and sequence, mention if the signal is isointense, hyperintense, etc.\\
For the impression:\\
1. Translate the following Chinese medical report impression into english.\\
2. List the diagnosed diseases mentioned within the english impression.\\
Here's the chinese report finding: <finding> and impression: <impression>. Please compile the results in the following JSON format:\{\\
    'finding':\{\\
        'english':'Translation of the chinese report fingding.',\
        'T1W': 'Description of the lesion in the original reports on T1W sequence, including its location, boundary, edema, its impact to surrouding brain tissue (if any), and the findings about midline, sulci, gyri, ventricles and cisterns.',\
        'T2W': 'Description of the lesion in the original reports on T1W sequence, including its location, boundary, edema, its impact to surrouding brain tissue (if any), and the findings about midline, sulci, gyri, ventricles and cisterns.',\
        'FLAIR': 'Description of the lesion in the original reports on T1W sequence, including its location, boundary, edema, its impact to surrouding brain tissue (if any), and the findings about midline, sulci, gyri, ventricles and cisterns.',\
        'T1C': 'Description of the lesion in the original reports on T1W sequence, including its location, boundary, edema, its impact to surrouding brain tissue (if any), and the findings about midline, sulci, gyri, ventricles and cisterns.'\\ 
        \},\\
    'impression':\{\\
        'translation': 'Translation of the chinese report impression.',\\
        'disease': [list of diseases mentioned in the english impression] \\
        \}\}
\end{mdframed}

\subsubsection{Prompt for ISLES2022 Report Processing}

\begin{mdframed}[backgroundcolor=gray!10]
GPT-4, I need you to analyze the following chinese MRI report fingdings and impression. For the finding:
    1. translate the provided report findings into english with precise medical terminology. Ensure that the translation accurately conveys the original meaning and includes all the relevant medical terms.\\
    2. Break the english fingdings down into a structured format. The JSON should organize the findings by anatomical regions and MRI sequences, with each related observation described in a complete sentence derieved from the original report. For each region and sequence, mention if the signal is isointense, hyperintense, etc.
    For the impression:
    1. Translate the following Chinese medical report impression into english.
    2. List the diagnosed diseases mentioned within the english impression.\\
    Here's the chinese report finding: <finding> and impression: <impression>. Please compile the results in the following JSON format:\\
    \{'finding':\{\\
        'english':'Translation of the chinese report fingding.',\\
        'ADC': 'Description of the signal of the lesion on ADC sequence (if any), and the findings about midline, sulci, gyri, ventricles and cisterns.',\\
        'DWI': 'Description of the signal of the lesion on DWI sequence (if any), and the findings about midline, sulci, gyri, ventricles and cisterns.'
    \},\\
    'impression':\{\\
        'translation': 'Translation of the chinese report impression.',\
        'disease': [list of diseases mentioned in the english impression]\
    \}
    \}\\
\end{mdframed}

\subsubsection{Prompt for WMH Report Processing}

\begin{mdframed}[backgroundcolor=gray!10]
GPT-4, I need you to analyze the following chinese MRI report fingdings and impression. For the finding:\\
    1. translate the provided report findings into english with precise medical terminology. Ensure that the translation accurately conveys the original meaning and includes all the relevant medical terms.\\
    2. Break the english fingdings down into a structured format. The JSON should organize the findings by anatomical regions and MRI sequences, with each related observation described in a complete sentence derieved from the original report. For each region and sequence, mention if the signal is isointense, hyperintense, etc.\\
    For the impression:\\
    1. Translate the following Chinese medical report impression into english.\\
    2. List the diagnosed diseases mentioned within the english impression.\\
    Here's the chinese report finding:'" + finding+"' and impression: <impression>. Please compile the results in the following JSON format:\\
    \{'finding':\{\\
        'english':'Translation of the chinese report fingding.',\\
        'T1W': 'Description of the signal of the lesion on T1W sequence (if any), and the findings about midline, sulci, gyri, ventricles and cisterns.',\\
        'FLAIR': 'Description of the signal of the lesion on FLAIR sequence (if any), and the findings about midline, sulci, gyri, ventricles and cisterns.'\\
    \},\\
    'impression':\{\\
        'translation': 'Translation of the chinese report impression.',\\
        'disease': [list of diseases mentioned in the english impression]\\
    \}\\
    \}
\end{mdframed}

\subsection{Human Evaluation Annotation Page}

\begin{figure}[!htb]
\centering
\includegraphics[width=\textwidth]{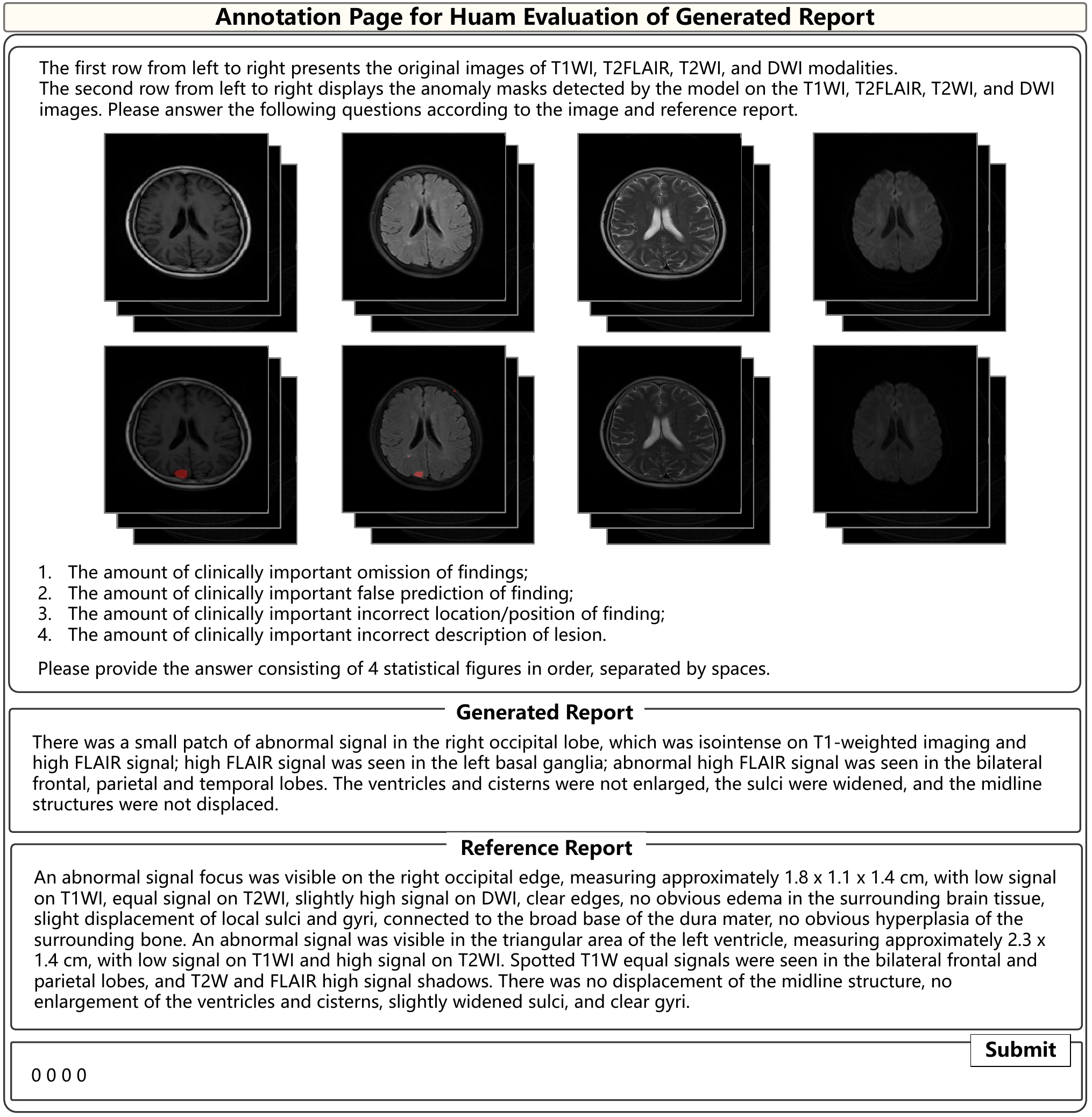}
\caption{The annotation page for human evaluation of the generated report.}
\label{fig:human_evaluation_annotation}
\end{figure}

\clearpage

\subsection{Examples for Generated Global Reports of \OURMODELWOS-Global/AutoSeg/Prompt}\label{sec:RG_three_setting}
\begin{figure}[!htb]
\centering
\includegraphics[width=\textwidth]{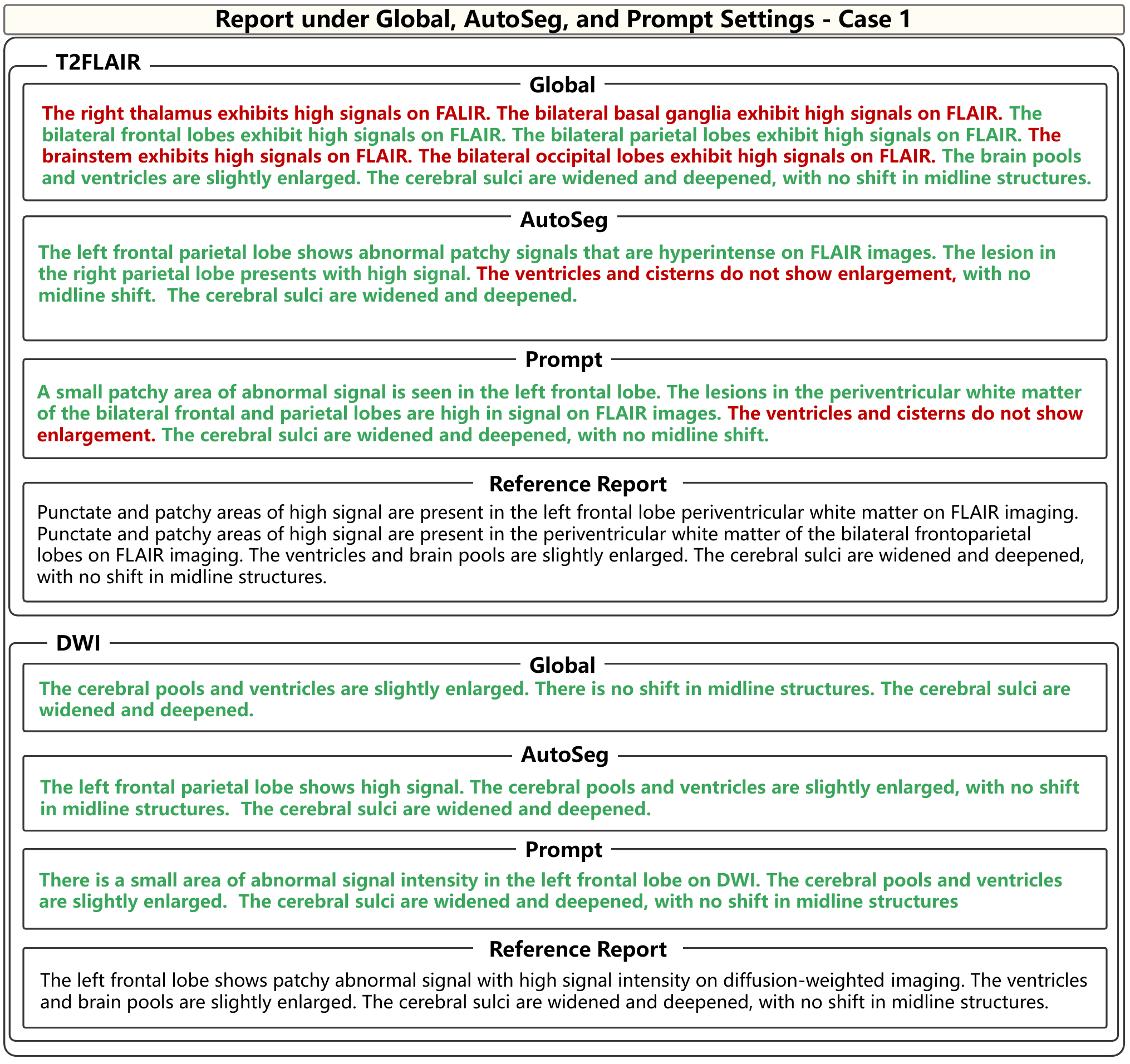}
\caption{The generated reports under three settings: \OURMODELWOS-Global, \OURMODELWOS-AutoSeg, and \OURMODELWOS-Promt, respectively. Red sentences are the incorrect findings. Green sentences are the correct findings.}
\label{fig:report_case1}
\end{figure}

\begin{figure}[!htb]
\centering
\includegraphics[width=\textwidth]{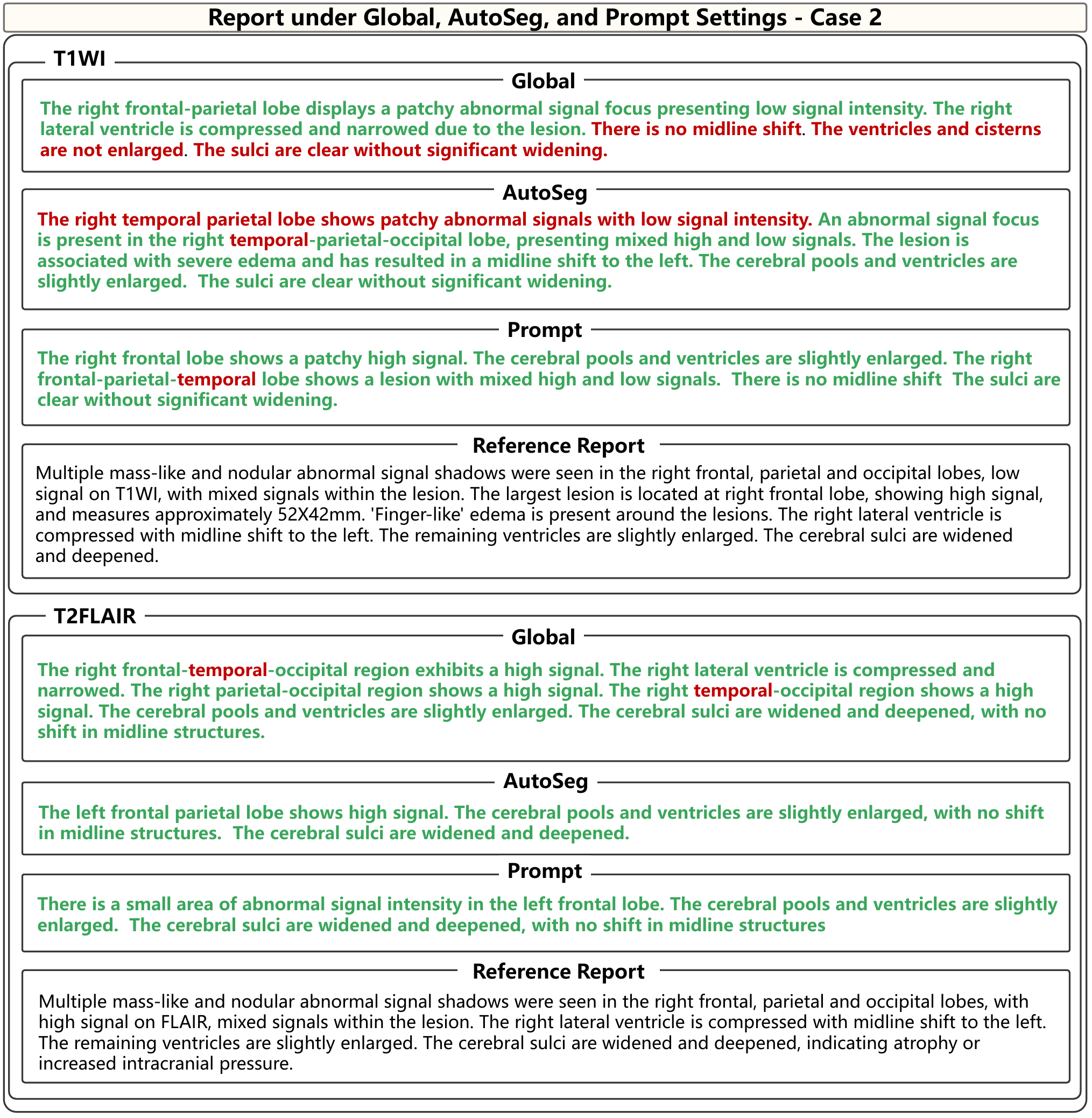}
\caption{The generated reports under three settings: \OURMODELWOS-Global, \OURMODELWOS-AutoSeg, and \OURMODELWOS-Promt, respectively. Red sentences are the incorrect findings. Green sentences are the correct findings.}
\label{fig:report_case2}
\end{figure}

\begin{figure}[!htb]
\centering
\includegraphics[width=\textwidth]{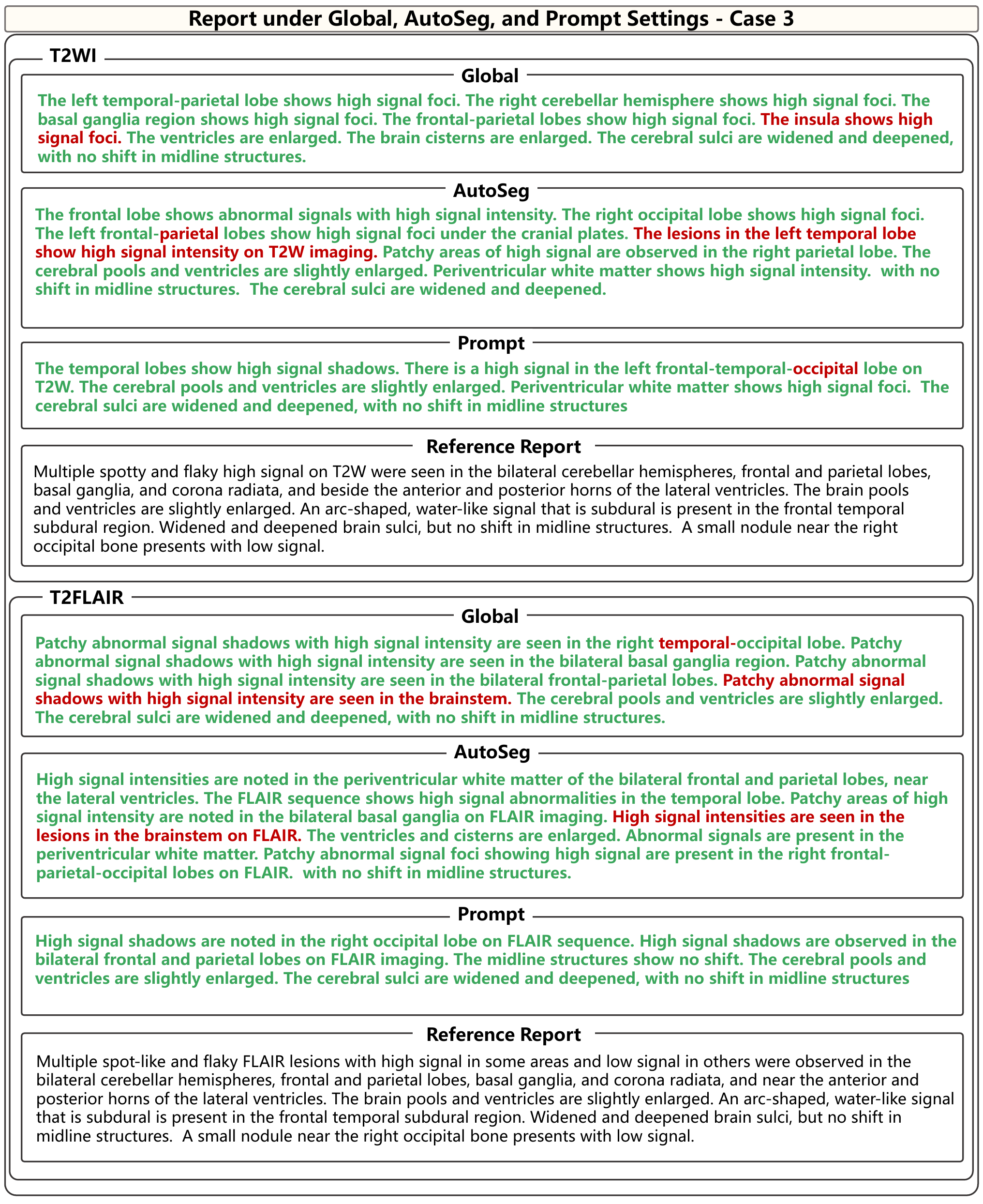}
\caption{The generated reports under three settings: \OURMODELWOS-Global, \OURMODELWOS-AutoSeg, and \OURMODELWOS-Promt, respectively. Red sentences are the incorrect findings. Green sentences are the correct findings.}
\label{fig:report_case3}
\end{figure}
\clearpage

\section{AI-assisted Report Writing}

\subsection{Annotation Page}

\begin{figure}[!htb]
\centering
\includegraphics[width=\textwidth]{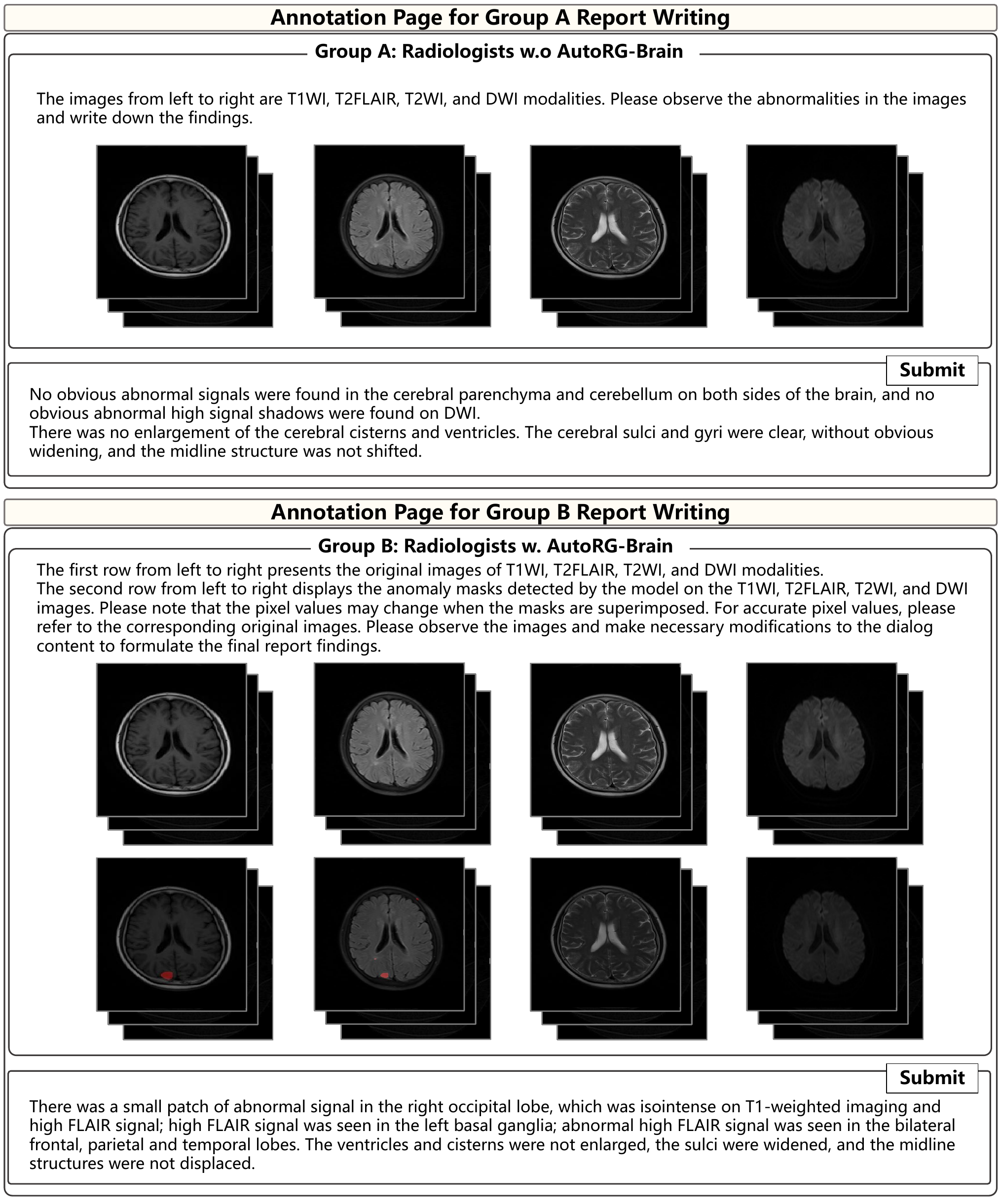}
\caption{The annotation page for doctors with and without AI assistance.}
\label{fig:report_writing_annotation}
\end{figure}

\clearpage

\subsection{Quantitative Results}

\begin{table}[!htb]
\centering
\footnotesize
\tabcolsep=0.1cm
   \vspace{10pt}
    \caption{The comparison results between the ground truth report and reports written by \OURMODEL, doctors w.o. \OURMODEL~(A1, A2, and A3), and doctors w. \OURMODEL~(B1, B2, and B3). BLEU-3, BLEU-4, ROUGE-1, Bert-Score, RadGraph, RadCliQ, and RaTEScore are reported.}
    \label{tab:human_accuracy}
    \setlength{\tabcolsep}{5pt}
   \begin{tabular}{c|ccccccc}
    \toprule
    Method & BLEU-3 $\uparrow$ & BLEU-4 $\uparrow$ & ROUGE-1 $\uparrow$ & Bert-Score $\uparrow$ & RadGraph $\uparrow$ & RadCliQ $\downarrow$ & RaTEScore $\uparrow$ \\ 
    \midrule
    \rowcolor{mygray}\multicolumn{8}{c}{\OURMODEL}\\
    \OURMODEL & 10.09$\pm$8.16 & 6.23$\pm$7.10 & 38.78$\pm$11.72 & 73.38$\pm$4.75 & 35.45$\pm$11.50 & 0.54$\pm$0.36 & 62.09$\pm$9.36 \\
    \midrule
    \rowcolor{mygray}\multicolumn{8}{c}{Human}\\
    A1 & 15.57$\pm$10.52 & 10.41$\pm$9.78 & 50.68$\pm$10.85 & 76.66$\pm$4.12 & 38.92$\pm$10.31 & 0.34$\pm$0.32 & 61.25$\pm$7.10 \\
    A2 & 11.76$\pm$9.99 & 7.47$\pm$9.21 & 42.52$\pm$9.92 & 74.31$\pm$4.24 & 34.87$\pm$9.88 & 0.52$\pm$0.33 & 57.54$\pm$6.32 \\
    A3 & 11.79$\pm$8.99 & 7.61$\pm$8.29 & 45.08$\pm$9.81 & 74.23$\pm$3.88 & 33.08$\pm$9.24 & 0.56$\pm$0.32 & 59.57$\pm$7.25\\
    \midrule
    Avg & 13.04$\pm$10.01 & 8.49$\pm$9.21 & \textbf{46.09$\pm$10.76} & 75.07$\pm$4.23 & 35.62$\pm$10.12 & 0.47$\pm$0.34 & 59.46$\pm$7.07\\
    \midrule
    \rowcolor{mygray}\multicolumn{8}{c}{Human w. \OURMODEL}\\
    B1 & 11.73$\pm$10.01 & 7.28$\pm$9.20 & 44.71$\pm$11.95 & 76.23$\pm$4.54 & 40.49$\pm$11.95 & 0.29$\pm$0.35 & 64.95$\pm$9.26\\
    B2 & 12.80$\pm$8.37 & 8.41$\pm$7.42 & 44.00$\pm$9.59 & 76.52$\pm$4.09 & 38.81$\pm$10.24 & 0.33$\pm$0.30 & 64.33$\pm$9.44\\
    B3 & 17.05$\pm$13.25 & 12.57$\pm$12.04 & 45.22$\pm$14.63 & 77.67$\pm$6.04 & 44.55$\pm$15.03 & 0.24$\pm$0.45 & 67.66$\pm$11.58\\
    \midrule
    Avg & \textbf{13.86$\pm$10.98} & \textbf{9.42$\pm$10.00} & 44.65$\pm$12.24 & \textbf{76.81$\pm$5.00} & \textbf{41.28$\pm$12.79} & \textbf{0.29$\pm$0.38} & \textbf{65.64$\pm$10.25}\\
    \bottomrule
    \end{tabular}
  \vspace{4pt}
\end{table}
\clearpage

\subsection{Case Study}\label{sec:case_ai_assisted}

\begin{figure}[!htb]
\centering
\includegraphics[width=\textwidth]{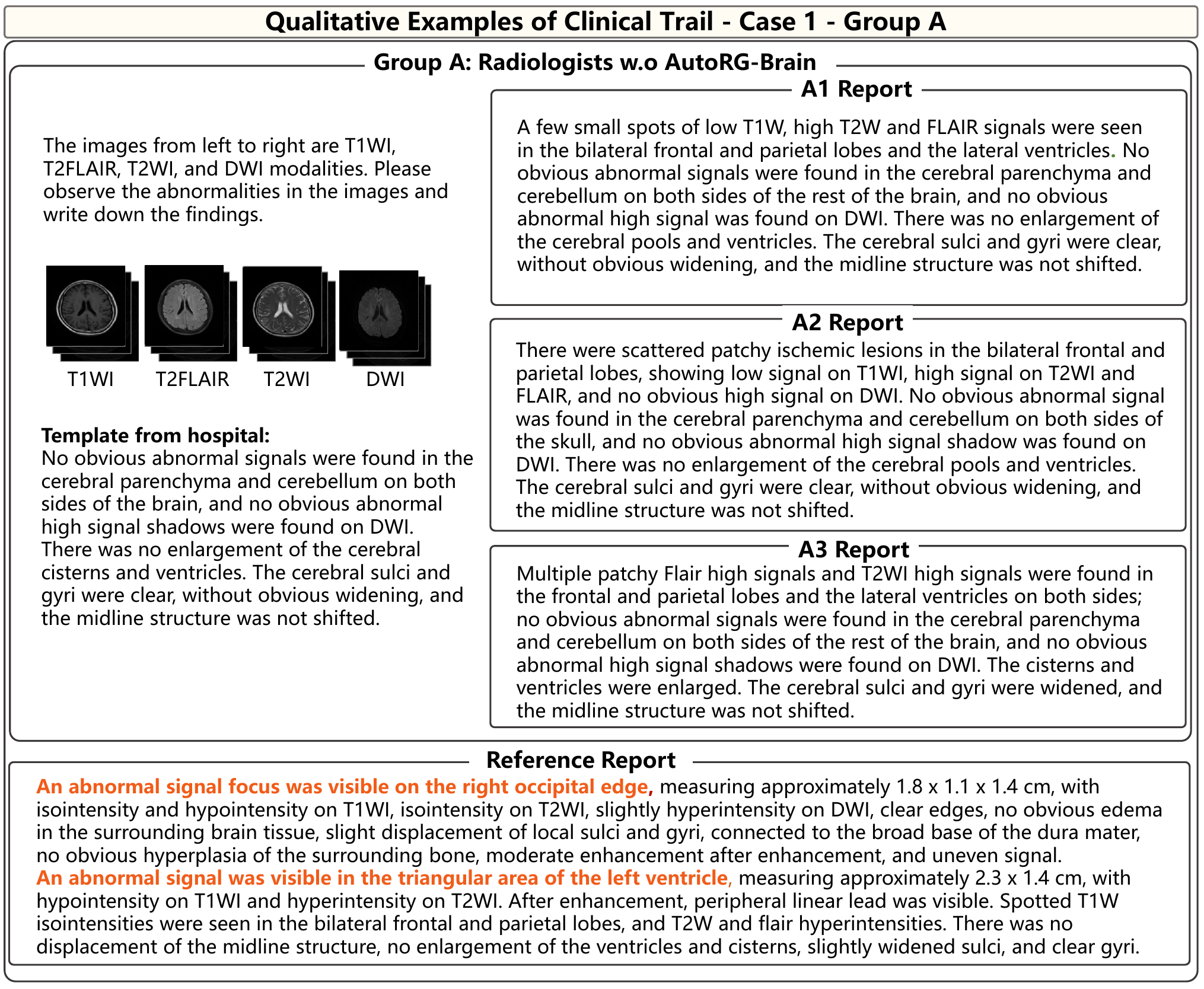}
\caption{The report written by three radiologists without AI-assistance in group A. Orange sentences are the findings the radiologists missed.}
\label{fig:Case_Study_1_GroupA}
\end{figure}

\begin{figure}[!htb]
\centering
\includegraphics[width=\textwidth]{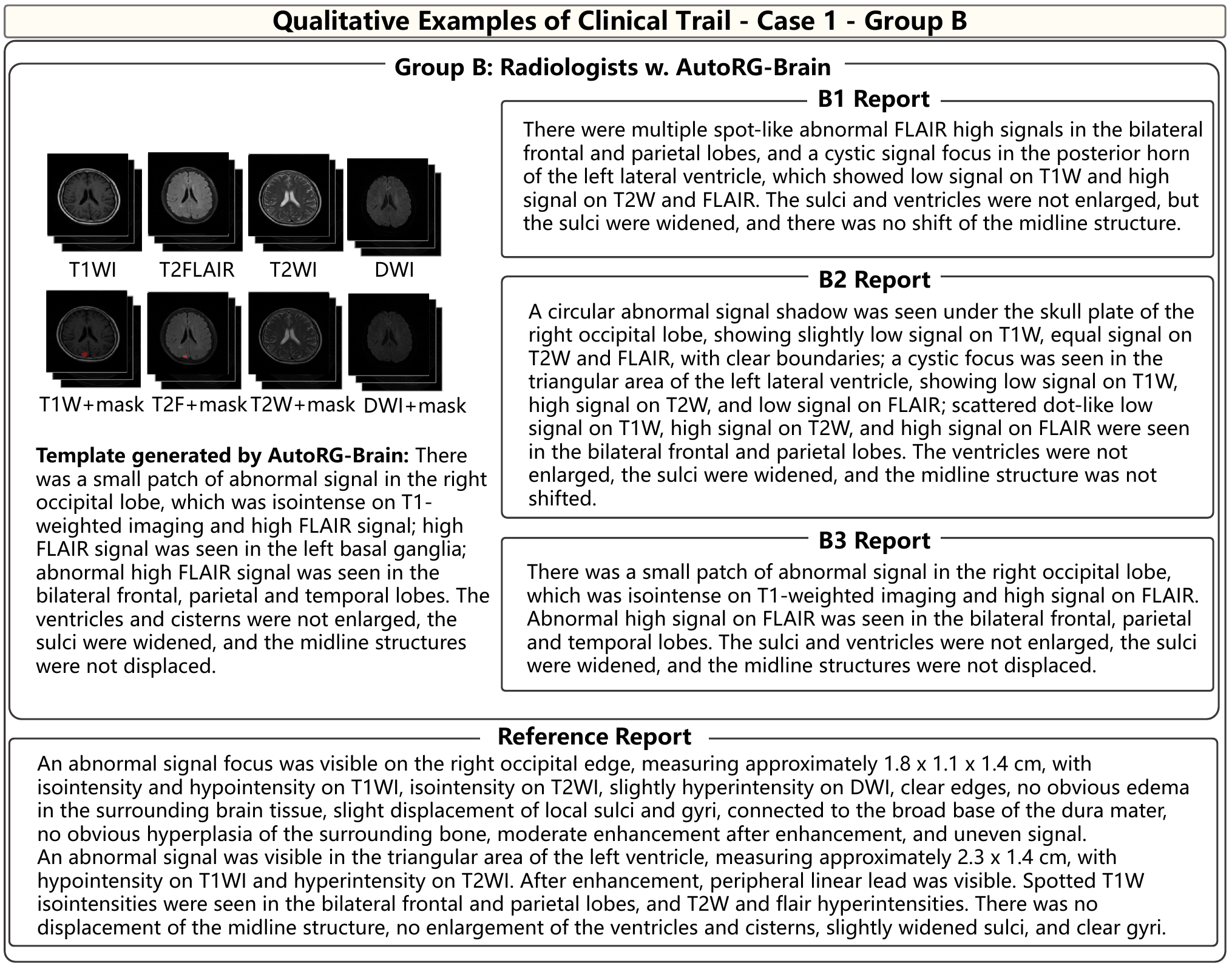}
\caption{The report written by three radiologists with AI-assistance in group B.}
\label{fig:Case_Study_1_GroupB}
\end{figure}

\begin{figure}[!htb]
\centering
\includegraphics[width=\textwidth]{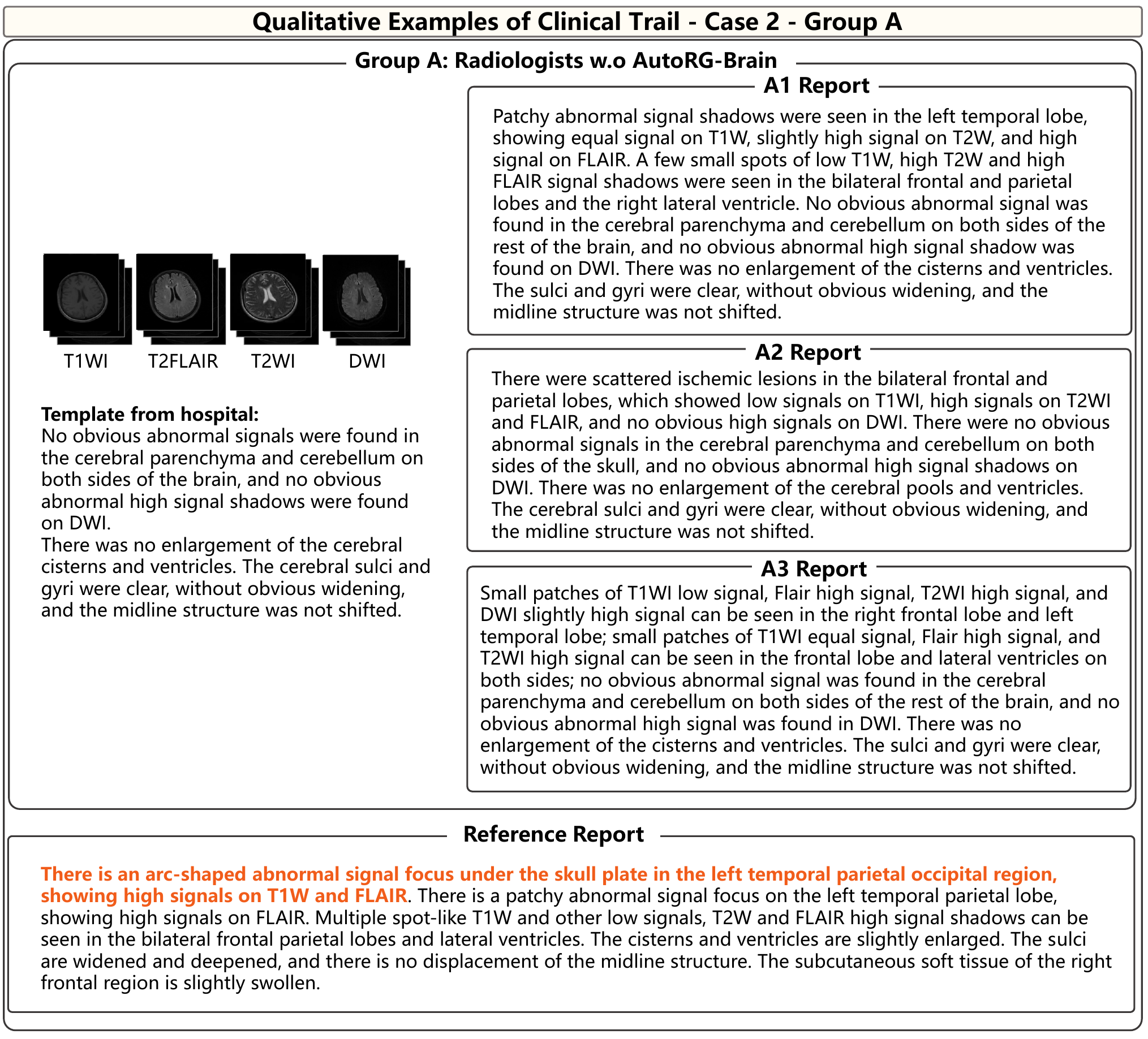}
\caption{The report written by three radiologists without AI-assistance in group A. Orange sentences are the findings the radiologists missed.}
\label{fig:Case_Study_2_GroupA}
\end{figure}

\begin{figure}[!htb]
\centering
\includegraphics[width=\textwidth]{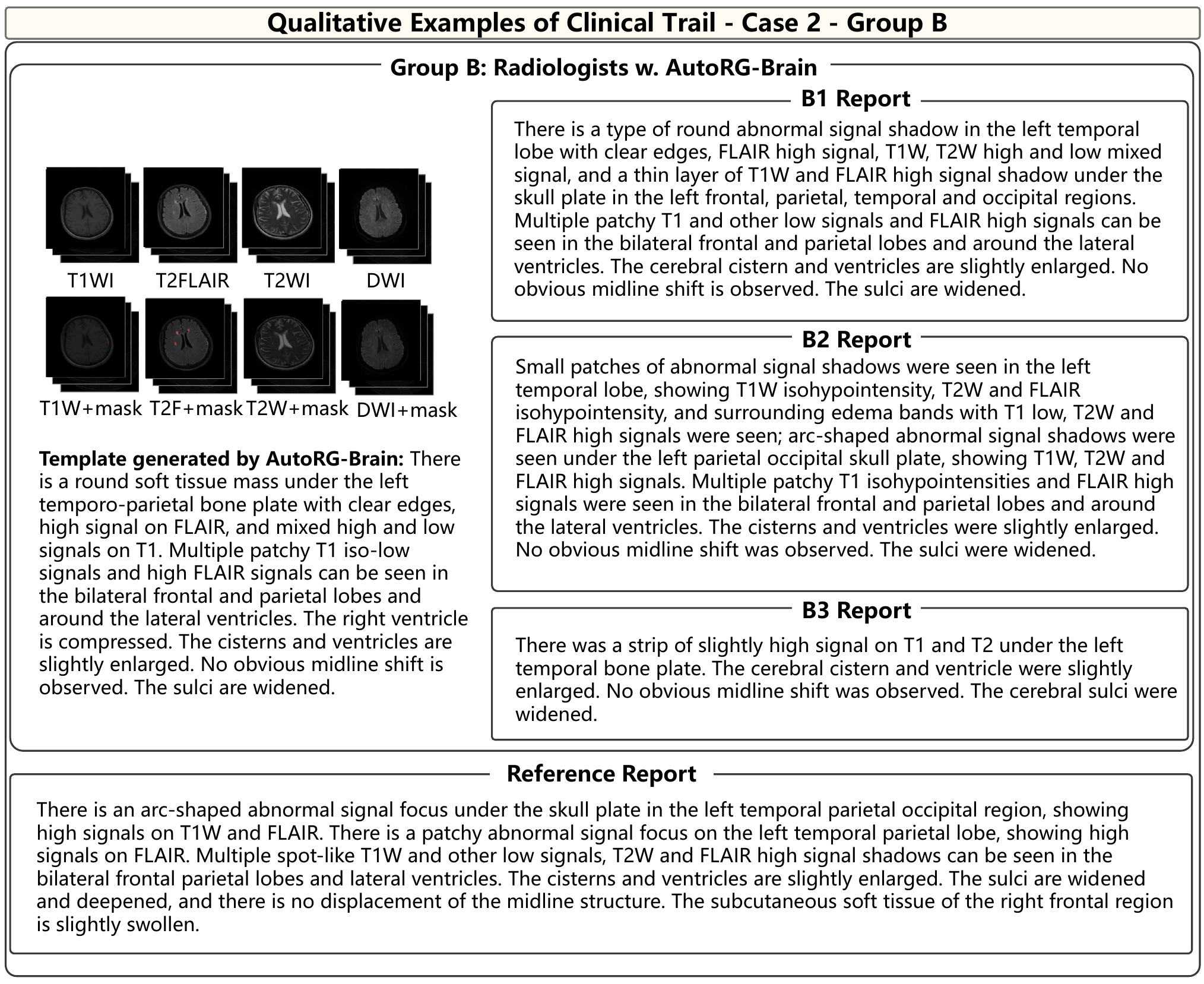}
\caption{The report written by three radiologists with AI-assistance in group B.}
\label{fig:Case_Study_2_GroupB}
\end{figure}

\begin{figure}[!htb]
\centering
\includegraphics[width=\textwidth]{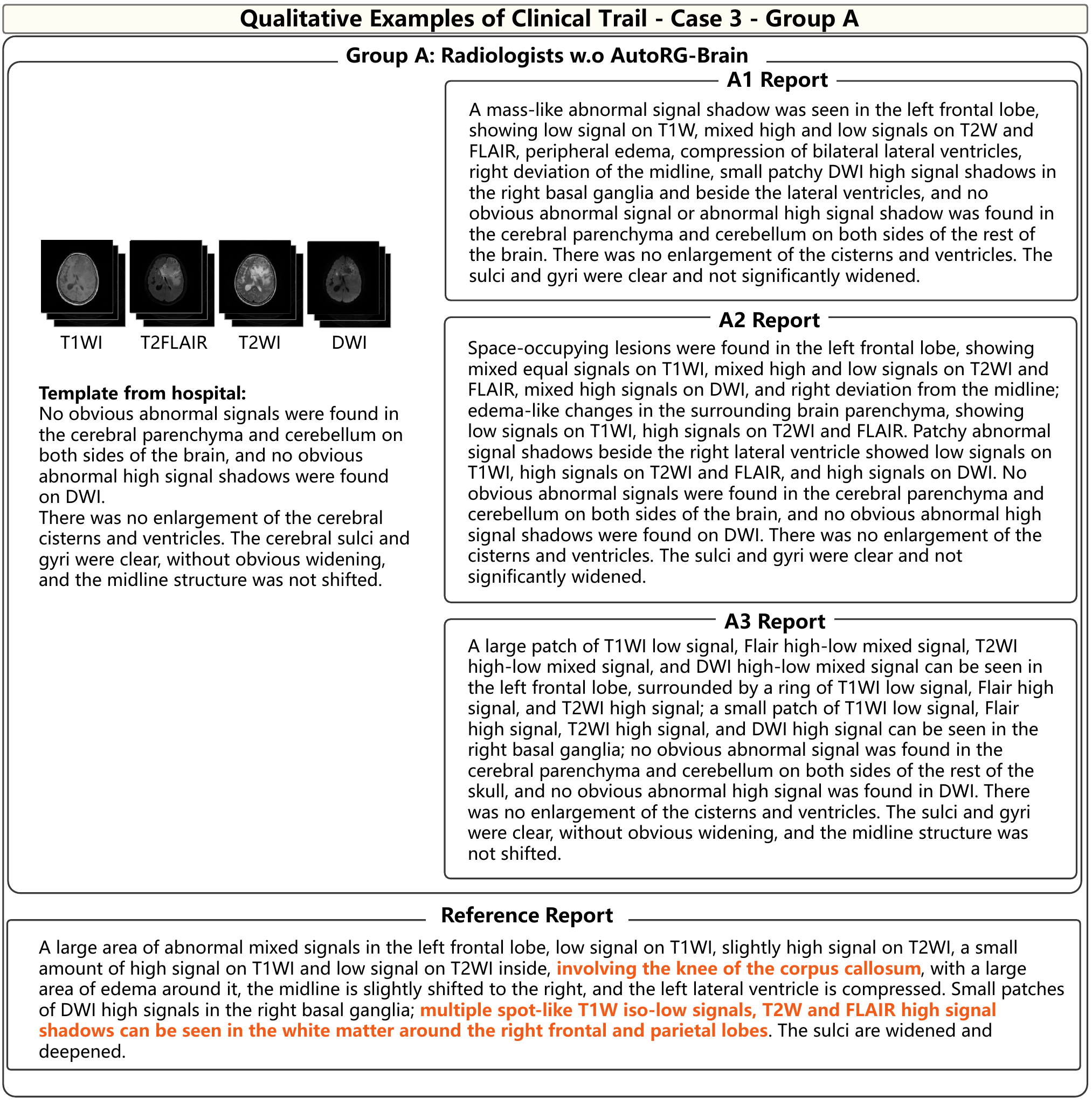}
\caption{The report written by three radiologists without AI-assistance in group A. Orange sentences are the findings the radiologists missed.}
\label{fig:Case_Study_3_GroupA}
\end{figure}

\begin{figure}[!htb]
\centering
\includegraphics[width=\textwidth]{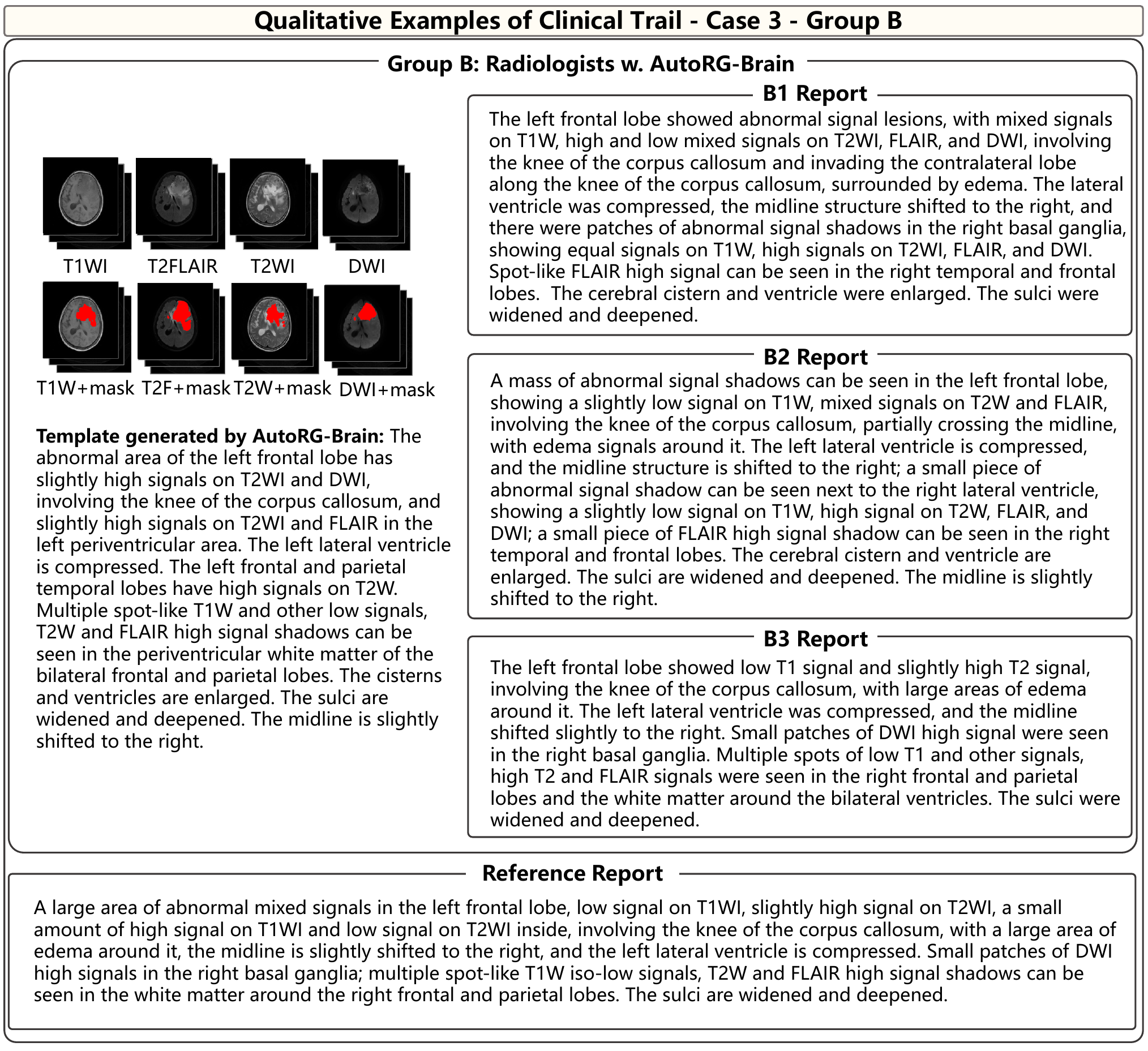}
\caption{The report written by three radiologists with AI-assistance in group B.}
\label{fig:Case_Study_3_GroupB}
\end{figure}

\end{document}